\documentclass[usenatbib]{mnras}

\usepackage{graphicx}
\usepackage{graphics}
\usepackage{amsmath}
\usepackage{amssymb}
\usepackage{color}
\usepackage{url}
\usepackage{hyperref}
\usepackage{widetext}

\def\aap{A\&A}
\def\aj{AJ}
\def\apj{ApJ}
\def\apjl{ApJL}
\def\apjs{ApJS}

\def\mnras{MNRAS}
\def\nat{Nature}
\def\aapr{A\&A Rev.}
\def\aaps{A\&AS}

\begin{document}

\title[2nd Generation Stars from Radiatively Cooled Winds]{Second Generation Stars in Globular Clusters from Rapid Radiative Cooling of Pre-Supernova Massive Star Winds}
\author[Lochhaas \& Thompson]{Cassandra Lochhaas$^{1}$ and Todd A. Thompson$^{1}$ \\ \\
$^{1}$ Department of Astronomy and the Center for Cosmology and Astroparticle Physics, The Ohio State University, \\
140 West 18th Avenue, Columbus, OH 43210, USA}

\maketitle

\begin{abstract}

Following work by W\"unsch and collaborators, we investigate a self-enrichment scenario for second generation star formation in globular clusters wherein wind material from first generation massive stars rapidly radiatively cools. Radiative energy loss allows retention of fast winds within the central regions of clusters, where it fuels star formation. Secondary star formation occurs in $\sim3-5$ Myr, before supernovae, producing uniform iron abundances in both populations. We derive the critical criteria for radiative cooling of massive star winds and the second generation mass as a function of cluster mass, radius, and metallicity. We derive a critical condition on $M/R$, above which second generation star formation can occur. We speculate that above this threshold the strong decrease in the cluster wind energy and momentum allows ambient gas to remain from the cluster formation process. We reproduce large observed second generation fractions of $\sim30-80\%$ if wind material mixes with ambient gas. Importantly, the mass of ambient gas required is only of order the first generation's stellar mass. Second generation helium enrichment $\Delta Y$ is inversely proportional to mass fraction in the second generation; a large second generation can form with $\Delta Y\sim0.001-0.02$, while a small second generation can reach $\Delta Y\sim0.16$. Like other self-enrichment models for the second generation, we are not able to simultaneously account for both the full range of the Na-O anticorrelation and the second generation fraction.
\end{abstract}

\begin{keywords}
galaxies: star clusters: general --- galaxies: star formation --- stars: winds, outflows
\end{keywords}

%%%%%%%%%%%%%%%%%%%%%%%%%%%%%%%%%%%%%%%%%%%%%%%%%%%%%%%%%%%%%%%%%%%%%%%%%%%%%%%%%%%%%%%%%
\section{Introduction}
%%%%%%%%%%%%%%%%%%%%%%%%%%%%%%%%%%%%%%%%%%%%%%%%%%%%%%%%%%%%%%%%%%%%%%%%%%%%%%%%%%%%%%%%%

Observations of globular clusters (GCs) imply that they contain more than a single simple stellar population. Star-to-star light element abundance variations, such as the O-Na and Mg-Al anticorrelations in red giants \citep{Gratton2001,Carretta2009a}, multiple main sequences \citep{Bedin2004,Piotto2007}, horizontal branches \citep{Ferraro1998,DAntona2002} and subgiant branches \citep{Bedin2004,Villanova2007}, and helium enrichment \citep{DAntona2005,Piotto2005} all indicate that each GC has at least two, and sometimes several, unique populations of stars. In most cases where just two stellar populations are identified, the so-called ``second generation" stars are $0.5-3$ times as prevalent as the ``first generation" \citep{Carretta2009a,Milone2017}.

Proposed ideas for the evolution of GCs containing multiple stellar populations include accretion of interstellar matter after the first star formation episode \citep{Bekki2009}, cluster mergers \citep{vandenBergh1996}, and several self-enrichment scenarios, in which ejecta from a first stellar generation fuels a second star formation episode. Light-element-enriched material is the result of hot H-burning mixed up to the convective zone of stars \citep{Denisenkov1990}. This material can then be expelled as winds from fast rotating stars \citep{Decressin2007a,Decressin2007b}, asymptotic giant branch (AGB) stars \citep{Ventura2001,Conroy2012}, supermassive stars \citep{Denissenkov2014}, massive binary stars \citep{deMink2009} or normal massive stars and supernovae \citep{Maeder2006,Prantzos2006,Tenorio2007,Wunsch2007,Wunsch2017}.

All proposed scenarios have problems explaining some aspects of the observations \citep[for recent reviews, see][]{Bastian2015b,Gratton2012}. In particular, there are four key issues facing all self-enrichment scenarios. First, the first-generation stars must supply enough material to form a massive second generation, which may include mixing with ambient gas. Second, the light element abundances of the second generation stars must match observations. These two points are closely tied, as the relative amounts of wind material and ambient gas affects both the abundances and size of the second generation. In order to produce second generation abundances matching observations from AGB or massive star wind material, GCs need to either have first stellar generations $10-100$ times more massive at birth and have significantly decreased their total stellar mass by ejecting most of the first generation stars into the Milky Way halo  \citep{Martell2011,Schaerer2011,Carretta2016}, or have a very top-heavy initial mass function (IMF) \citep{Conroy2012,Decressin2007b}. The third key issue is that wind speeds near or exceeding the escape velocity of GCs --- e.g., $10-30$ km s$^{-1}$ for AGB winds \citep{Loup1993} or $1000-2000$ km s$^{-1}$ for massive star winds \citep{Lamers1995} --- make it difficult for the shallow gravitational potential wells of GCs to retain wind material. Fourth, the second generation must form without supernova ejecta, since the stars in GCs have uniform iron abundances (see review by \citeauthor{Suntzeff1993} \citeyear{Suntzeff1993} and references therein, \citeauthor{Carretta2009b} \citeyear{Carretta2009b}; exceptions are $\omega$ Cen, \citeauthor{Gratton1982} \citeyear{Gratton1982} and \citeauthor{Johnson2010} \citeyear{Johnson2010}, and others, \citeauthor{Johnson2015} \citeyear{Johnson2015}).

We investigate a scenario for second generation formation explored by \citet{Wunsch2008,Palous2014,Wunsch2017} in which massive star winds from the first stellar generation shock and thermalize to produce a region of hot gas in the cluster interior on Myr timescales after the first generation's formation. If the mass loss rate is high enough or if the winds mix with ambient gas left over from cluster formation, the gas is dense enough to become radiative, loses its thermal energy, and can be retained in the GC to fuel a subsequent generation of star formation. The action of rapid radiative cooling eliminates the wind retention problem. We show that in order for radiative cooling to set in and form a second stellar generation $0.5-3$ times as massive as the first, wind mass-loading factors must be $\sim20-100$. This is achievable if the wind ejecta mixes with an ambient gas mass of order the first generation stellar mass, less if wind mass loss rates are higher than current stellar evolution models suggest. The ambient gas may be left over from the first stellar formation episode. In our picture, the GC does not need to have a larger first generation at birth than observed in GCs today, does not need a top-heavy IMF, and does not need to eject first generation stars. Finally, second generation formation occurs only within the first few Myr after the GC's birth, before supernovae from the most massive stars occur, so that there is no additional Fe enrichment in the second generation. Clusters likely blow out the remnants of their natal gas cloud $\sim5$ Myr after formation \citep{Bastian2014}, halting any additional second generation formation.

In section~\ref{sec:model}, we estimate a critical condition for cooling of gas deposited by massive star winds forming a second stellar generation. We make predictions for the required mass-loading factor and size of the second generation as functions of the GC's mass and radius, derive a critical condition on the cluster's stellar mass per unit radius $M/R$ for cooling to set in, compare to observations of $M/R$ and second-generation stellar fraction in GCs, and compare the observed helium and other light element abundance spreads to abundances in our model. In section~\ref{sec:discussion}, we discuss outstanding problems, uncertainties, and how changes to our assumptions impact our model. Section~\ref{sec:conclusions} gives our conclusions. 

%%%%%%%%%%%%%%%%%%%%%%%%%%%%%%%%%%%%%%%%%%%%%%%%%%%%%%%%%%%%%%%%%%%%%%%%%%%%%%%%%%%%%%%%%
\section{The Model}
\label{sec:model}
\subsection{Critical Condition for Cooling}
\label{sec:crit}
%%%%%%%%%%%%%%%%%%%%%%%%%%%%%%%%%%%%%%%%%%%%%%%%%%%%%%%%%%%%%%%%%%%%%%%%%%%%%%%%%%%%%%%%%

We wish to estimate the size of the region within the cluster that radiatively cools as a function of cluster properties. For low enough densities, the material should never be radiative and the thermalized winds will drive a fast cluster wind \citep{Silich2003,Silich2004}. For higher densities, radiative cooling should first set in at the cluster center, while the outer lower density regions continue to drive an outflow \citep{Tenorio2005,Tenorio2007,Wunsch2007,Wunsch2008}. Finally, for higher density, a substantial fraction of the cluster's gas will cool and self-shield \citep{Palous2014,Wunsch2017}, leading to star formation. We assume spherical symmetry, uniform mass and energy deposition, and no gravity. If the massive star wind ejecta with total kinetic power $\dot E_\mathrm{tot}$ thermalizes, the asymptotic cluster wind velocity from energy conservation is
\begin{align}
\frac{1}{2}\dot M_\mathrm{tot} v_\infty^2 &= \dot E_\mathrm{tot}\ \ \ \ \Longrightarrow\ \ \ \ v_\infty &= \left(2\frac{\alpha}{\beta}\frac{\dot E_w}{\dot M_w}\right)^{\frac{1}{2}} \label{eq:vinfty}
\end{align}
where $\dot M_w$ is the mass deposition rate from massive star winds, $\dot M_\mathrm{tot}$ is the \emph{total} mass deposition rate, which may include wind material and swept-up ambient cluster gas, and the mass-loading factor $\beta$ is the ratio between the two, such that $\dot M_\mathrm{tot}=\beta \dot M_w$. $\beta$ represents any additional mass deposition other than the $\dot M_w$ defined in equation~(\ref{eq:Mdot}), which may be ambient gas or additional stellar wind mass loss. $\dot E_w$ is the wind energy deposition, $\alpha$ is the efficiency of thermalization such that the total energy deposition is $\dot E_\mathrm{tot}=\alpha \dot E_w$. In the absence of radiative cooling, the hot, thermalized gas inside the cluster will expand out from the edge of the cluster, driving a supersonic wind. The sonic point will be located at the edge of the energy and mass deposition region at $R_{cl}$ \citep{CC1985,Wang1995,Silich2004}. Using the fact that the Bernoulli integral is constant for $r>R_{cl}$ and that the sonic point is at $R_{cl}$, $c_s^2=v_\infty^2/4$ at $r = R_{cl}$ for $\gamma=5/3$, where $v_\infty^2$ is given in equation~(\ref{eq:vinfty}). This allows us to write the temperature and pressure in terms of $\alpha$, $\beta$, $\dot E_w$, and $\dot M_w$. Given mass conservation, we can also find the density at $R_{cl}$.

The conditions for radiative cooling vary within the cluster because the density and temperature vary with position, and depend on the radial distribution of $\dot E_\mathrm{tot}$ and $\dot M_\mathrm{tot}$ per volume. For constant volumetric energy and mass injection rates, \citet{CC1985} derived the self-similar solution for pressure, density, and wind velocity. We use values from \citet{CC1985} to scale the density and temperature at $R_{cl}$ to values within the cluster: $T(r<R_{cl})\approx\zeta_T T(R_{cl})$ and $\rho(r<R_{cl})\approx\zeta_\rho \rho(R_{cl})$. Values of $\zeta_\rho=2.07$ and $\zeta_T=1.28$ are the averages of the scalings found in \citet{CC1985} between 10\% and 80\% of the cluster's volume (corresponding to radii within the cluster of $0.46R_{cl}$ and $0.93R_{cl}$). We use these as representative values throughout. For a different model of the radial dependence of the mass and energy injection rate per unit volume, $\zeta_\rho$ and $\zeta_T$ will change and may be functions of the radius within the cluster \citep[see Appendix of][]{Zhang2014}. Hydrodynamical calculations should assess the role of multidimensional effects like clumping \citep[e.g.,][]{Wunsch2017}.

As a reference point, we adopt values for $\dot E_w$ and $\dot M_w$ as the average values from 2.5 Myr to 7 Myr given by a combination of the codes STARBURST99 \citep{Leitherer1999,Leitherer2014} and BPASS \citep{Eldridge2008,Eldridge2009} for an instantaneously-formed $10^5\ M_\odot$ GC with a double power law initial mass function (IMF), with power law index for d$N$/d$M$ of $-1.3$ between $0.1M_\odot$ and $0.5M_\odot$ and $-2.35$ between $0.5M_\odot$ and $120M_\odot$, at one-tenth solar metallicity. Although the relation between initial mass and the probability of black hole formation is complicated \citep{Pejcha2015,Sukhbold2016}, we set the mass cutoff for direct collapse to black hole at $25M_\odot$, which means the first supernova can be expected at $\sim7$ Myr. Before $2.5$ Myr, the mass deposition rates are low enough to be negligible because stars have not yet entered post-main sequence evolution. We mix the two codes by assuming a 50\% binary fraction, and we have selected only those binaries in BPASS with mass ratio $0.9$ and period $10^4$ days. Other mass ratios and periods vary the STARBURST99 and BPASS combined average mass loss rates from $10^{-3.4}\ M_\odot$ yr$^{-1}$ to $10^{-3.1}\ M_\odot$ yr$^{-1}$, so we pick the period and mass ratio combination that maximizes $\dot M_w$. Shorter binary periods decrease the average $\dot M_w$ slightly. The BPASS code examines binary mass loss as a function of system mass, which is useful for constructing a GC with a specific IMF. However, \citet{deMink2009} proposes that massive binaries in a cluster produce a total wind mass that contains 13\% of the first generation stellar mass, implying a mass loss rate of $\sim10^{-2.5}\ M_\odot$ yr$^{-1}$, 0.6 dex higher than our fiducial mass loss rate, so the rates we use here may be somewhat conservative. The amount of wind ejecta relative to the amount of ambient gas swept up is crucial in determining the helium enrichment of the second generation. For this reason, we consider enhancements to our fiducial $\dot M_w$ in \S\ref{sec:He_abundances}, and we consider the maximum possible mass loss rate for a massive star population.

Only STARBURST99 provides energy deposition rates, and so we assume this $\dot E_w$ for the binary stars from BPASS. We further assume both $\dot E_w$ and $\dot M_w$ scale linearly with cluster mass above $10^4M_\odot$, as long as the IMF is fully populated. The average values\footnote{We find the average values of $\dot E_w$ and $\dot M_w$ by summing the total energy and mass deposition from both STARBURST99 and BPASS, then dividing by the total time from cluster formation to first supernova. BPASS examines single binary systems with discrete masses, not a full binary stellar population. We combine the BPASS systems into a full cluster by integrating with our IMF, but the system masses examined in BPASS sparsely sample the IMF at the high mass end. This produces a bursty total mass- and energy-loss rate over the full evolution of the cluster, but with a well-defined average value given in equations~(\ref{eq:Edot}) and~(\ref{eq:Mdot}). Including the BPASS code in our calculation increases $\dot M_w$ by 0.4 dex over just STARBURST99 for $Z=0.1Z_\odot$, and by 0.1 dex for $Z=0.7Z_\odot$.} for our fiducial cluster are then
\begin{align}
\dot E_w &= 10^{37.8}\ \mathrm{ergs\ s}^{-1}\ \left(\frac{M_1}{10^5\ M_\odot}\right) \label{eq:Edot} \\
\dot M_w &= 10^{-3.1}\ M_\odot\ \mathrm{yr}^{-1}\ \left(\frac{M_1}{10^5\ M_\odot}\right) \label{eq:Mdot}
\end{align}
where $M_1$ is the mass of the first generation of stars, which is the total stellar mass of the cluster before formation of the second stellar generation. For these values, $v_\infty\approx500$ km s$^{-1}$ (eq.~\ref{eq:vinfty}). We also explore an additional value for the cluster metallicity closer to solar for comparison. For metallicity $0.7Z_\odot$, $\dot E_w=10^{38.7}$ ergs s$^{-1}$ $(M_1/10^5M_\odot)$ and $\dot M_w=10^{-3.0}\ M_\odot$ yr$^{-1}$ $(M_1/10^5M_\odot)$, implying $v_\infty\approx1200$ km s$^{-1}$. Our fiducial cluster has a first generation stellar mass $M_1=10^5\ M_\odot$, half-light radius $R_{cl}=1$ pc, one-tenth solar metallicity, and efficiency of energy thermalization $\alpha=0.1$, but we explore different values throughout.

Because the advection time is larger than the heating time throughout most of the cluster volume \citep{CC1985}, we neglect the kinetic energy of the thermalized gas and assume the cluster cools radiatively if the volumetric cooling rate is larger than the volumetric heating rate $\dot E_\mathrm{tot}$: $\Gamma_\mathrm{cool}\geq \Gamma_\mathrm{heat}$. For the purposes of an analytic estimate, we approximate the cooling rate as a double power-law below $10^7$ K \citep{Draine2011},
\begin{equation}
\Gamma_\mathrm{cool}=\left\{
\begin{array}{lr}
\Lambda_0 \left(\frac{T_0}{T}\right)^{0.7} n^2 & 10^5<T<10^7\ \mathrm{K} \\
\Lambda'_0 \left(\frac{T}{T'_0}\right)^p n^2 & 10^4 < T < 10^5\ \mathrm{K}
\end{array}
\right. \label{eq:cooling}
\end{equation}
where $n$ and $T$ are the number density and temperature of the gas. At one-tenth solar metallicity, $\Lambda_0\simeq2.2\times10^{-23}$ ergs s$^{-1}$ cm$^3$, $T_0=10^6$ K, $\Lambda'_0\simeq1.1\times10^{-22}$ ergs s$^{-1}$ cm$^3$, $p=0.3$, and $T'_0=10^5$ K; at $0.7$ solar metallicity, $\Lambda_0\simeq9.0\times10^{-23}$ ergs s$^{-1}$ cm$^3$, $T_0=10^6$ K, $\Lambda'_0\simeq4.5\times10^{-22}$ ergs s$^{-1}$ cm$^3$, $p=0.9$, and $T'_0=10^5$ K. The local heating rate is
\begin{equation}
\Gamma_\mathrm{heat}=\frac{3\alpha \dot E_w}{4\pi R_{cl}^3}.
\end{equation}
We will focus on one-tenth solar metallicity and $T>10^5$ K, but the following derivation is similar, with only differing exponential powers, for different metallicities and temperatures. Setting the heating and cooling rates equal allows us to derive the critical condition for cooling:
\begin{equation}
\frac{3\alpha \dot E_w}{4\pi R_{cl}^3} = \Lambda_0\left(\frac{T_0}{T}\right)^{0.7} n^2 \label{eq:balance}
\end{equation} 
The density inside the thermalized region is $n\simeq\zeta_\rho n(R_{cl})$, where
\begin{equation}
n(R_\mathrm{cl})=\frac{\beta \dot M_\mathrm{esc}}{4\pi R_{cl}^2 \mu m_p c_s}, \label{eq:dens}
\end{equation}
where $\dot M_\mathrm{esc}$ is the mass deposition rate of only that material that escapes the cluster as a wind (the gas that does not cool radiatively). The temperature of thermalized gas is given by
\begin{equation}
T=\zeta_T \frac{3}{5} \frac{\mu m_p}{k_B} c_s(R_{cl})^2=\zeta_T \frac{3}{10} \frac{\mu m_p}{k_B}\frac{\alpha \dot E_w}{\beta \dot M_w}. \label{eq:temp}
\end{equation}
Scaling this to our fiducial cluster yields
\begin{equation}
T\sim 4\times10^5\ \mathrm{K}\ \frac{\alpha_{0.1}}{\beta_{2}}, \label{eq:temp_scaled}
\end{equation}
where $\alpha_{0.1}=\alpha/0.1$ and we have anticipated the mass loading factor needed for most of the cluster volume to cool (see eq.~\ref{eq:beta_crit_scaled}) and scaled to $\beta_{2}=\beta/2$.

\begin{figure}
\includegraphics[width=\linewidth]{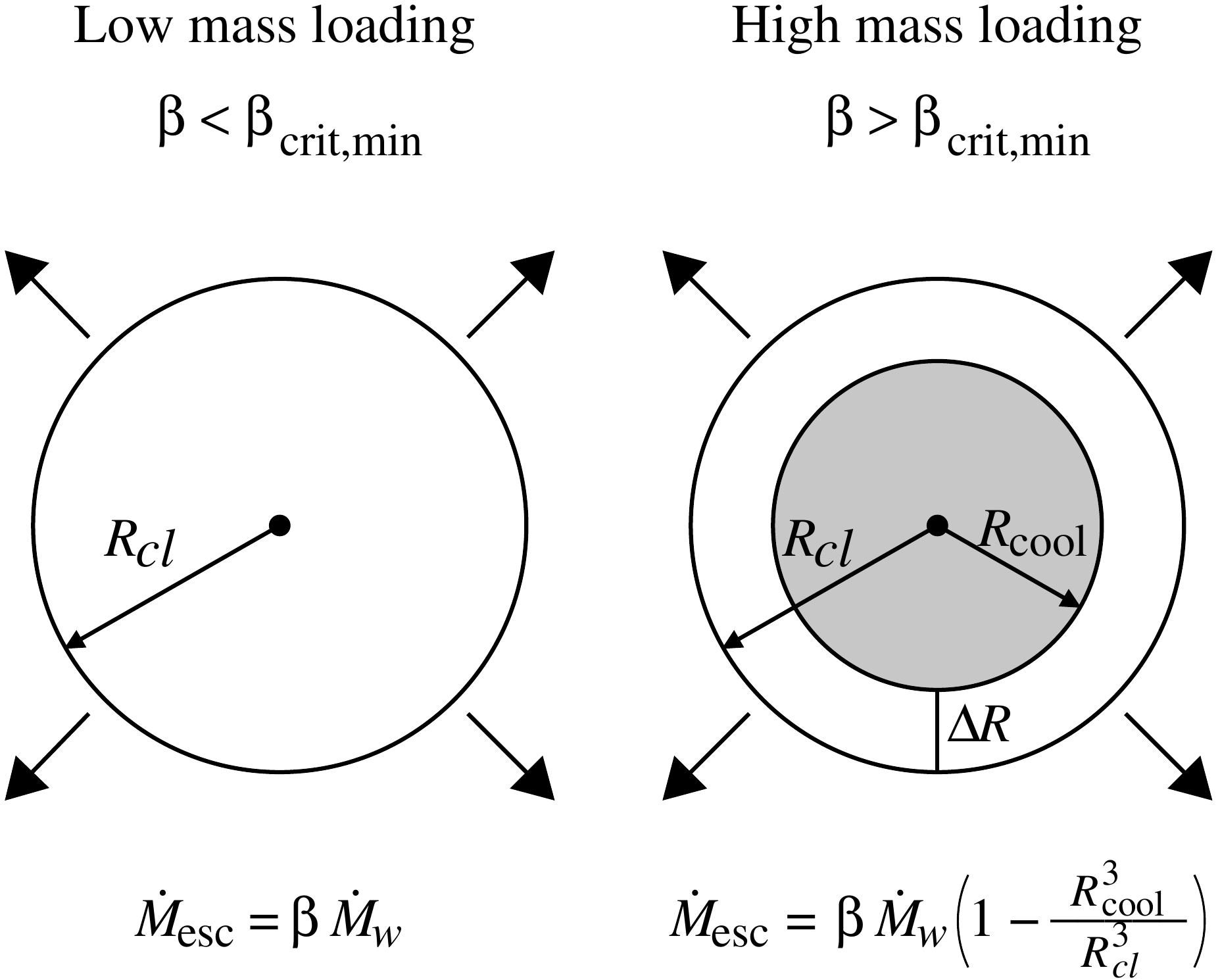}
\caption{A cartoon showing the two scenarios of massive star winds depositing energy and mass in a region of size $R_{cl}$. On the left, there is not enough mass loading of the winds ($\beta<\beta_\mathrm{crit,min}$) and all deposited mass exits the cluster as a wind. On the right, $\beta>\beta_\mathrm{crit,min}$ and the inner shaded region within $R_\mathrm{cool}$ (see equation~\ref{eq:Rcool}) cools and is retained in the cluster while any mass deposited in the outer region between $R_\mathrm{cool}$ and $R_{cl}$ escapes the cluster as a wind. The critical $\beta$ for a given $\Delta R$ is given in equation~(\ref{eq:beta_crit_scaled}).}
\label{fig:cartoon}
\end{figure}

We define a cooling radius $R_\mathrm{cool}$ such that all mass and energy deposited within this radius cools and stays within the cluster, whereas all mass and energy deposited outside $R_\mathrm{cool}$ escapes as a wind. Since we assume uniform mass and energy deposition everywhere within $R_{cl}$, $\dot M_\mathrm{esc}=\beta\dot M_\mathrm{tot} \left[1-\left(\frac{R_\mathrm{cool}}{R_{cl}}\right)^3\right]$ (see Figure~\ref{fig:cartoon}).

Substituting equations~(\ref{eq:dens}) and~(\ref{eq:temp}) into equation~(\ref{eq:balance}) and solving for $R_\mathrm{cool}$ yields
\begin{equation}
\left(\frac{R_\mathrm{cool}}{R_{cl}}\right)^3=1-\sqrt{\left(\frac{3\zeta_T}{10k_B T_0}\right)^{0.7}\frac{6\pi(\mu m_p)^{2.7} R_{cl}\alpha^{2.7}\dot E_w^{2.7}}{\Lambda_0\beta^{3.7}\dot M_w^{3.7}\zeta_\rho^2}}. \label{eq:Rcool}
\end{equation}
Solving for the critical $\beta$ value needed for a given $R_\mathrm{cool}$ yields
\begin{equation}
\beta_\mathrm{crit}^{3.7}=\left(\frac{3\zeta_T}{10k_BT_0}\right)^{0.7}\frac{6\pi (\mu m_p)^{2.7} R_{cl} \alpha^{2.7} \dot E_w^{2.7}}{\Lambda_0 \dot M_w^{3.7}\left[1-\left(\frac{R_\mathrm{cool}}{R_{cl}}\right)^3\right]^2\zeta_\rho^2}. \label{eq:beta_crit}
\end{equation}
In the limit that the cooling radius approaches $R_{cl}$, $\left[1-\left(\frac{R_\mathrm{cool}}{R_{cl}}\right)^3\right]^2\approx 9\left(\frac{\Delta R}{R_{cl}}\right)^2$, where $\Delta R=R_{cl}-R_\mathrm{cool}$. Scaling this limit to our fiducial globular cluster properties, we find
\begin{equation}
\beta_\mathrm{crit}\sim 2\ \alpha_{0.1}^{0.73} \left(\frac{R_{cl}}{\mathrm{pc}}\right)^{0.81} \left(\frac{10^5 M_\odot}{M_1}\right)^{0.27} \left(\frac{0.1\ \mathrm{pc}}{\Delta R}\right)^{0.54} \label{eq:beta_crit_scaled}
\end{equation}
for one-tenth solar metallicity gas and $\mu=1.38$. If $\beta\geq\beta_\mathrm{crit}$, the gas within $R_\mathrm{cool}$ radiatively cools and fuels the second generation of star formation. In the limit $R_\mathrm{cool}\rightarrow 0$, we find\footnote{We rearranged our equation for $\beta_\mathrm{crit,min}$ to instead solve for $\dot E_\mathrm{crit}$ and compared with $L_\mathrm{crit}$ from \citet{Wunsch2007}. For similar GC properties, we find our calculation and its scaling with GC parameters agree.}
\begin{equation}
\beta_\mathrm{crit,min}\sim1\ \alpha_{0.1}^{0.73} \left(\frac{R_{cl}}{\mathrm{pc}}\right)^{0.27}\left(\frac{10^5 M_\odot}{M_1}\right)^{0.27}. \label{eq:beta_crit_min}
\end{equation}
If $\beta<\beta_\mathrm{crit,min}$, none of the gas in the cluster cools, and no second generation of star formation occurs. Note that the critical value for $\beta$ for nearly all of the cluster volume to cool is just $\sim2\times\beta_{\mathrm{crit,min}}$. We can rearrange equation~(\ref{eq:beta_crit_min}) to find a condition on $M_1$ and $R_{cl}$ for cooling for a given $\alpha$ and $\beta$:
\begin{equation}
\left[\frac{M_1}{R_{cl}}\right]_\mathrm{crit,min}\sim 10^5\ M_\odot\ \mathrm{pc}^{-1}\ \alpha_{0.1}^{2.7}\beta^{-3.7}. \label{eq:M1dRcrit}
\end{equation}
If the combination of first-generation mass and cluster radius are not larger than $\left[\frac{M_1}{R_{cl}}\right]_\mathrm{crit,min}$ for a given $\alpha$ and $\beta$, then we would not expect a second generation of stars to form at all in that cluster. Similarly, we can rewrite equation~(\ref{eq:beta_crit_scaled}) to find a condition on $M_1$ and $R_{cl}$ for nearly all of the cluster to cool for a given $\alpha$ and $\beta$:
\begin{align}
\left[\frac{M_1}{R_{cl}}\right]_\mathrm{crit}&\sim 2.3\times10^5\ M_\odot\ \mathrm{pc}^{-1}\ \frac{\alpha_{0.1}^{0.9}}{\beta^{1.2}} \nonumber \\
&\times\left(\frac{M_1}{10^5\ M_\odot}\right)^{0.67}\left(\frac{0.1\ \mathrm{pc}}{\Delta R}\right)^{0.67}. \label{eq:M1dRcritfullcool}
\end{align}

Equations~(\ref{eq:beta_crit_scaled}) and~(\ref{eq:beta_crit_min}) imply that for a significant fraction of the cluster's gas to cool, $\dot M_w$ must be of order the fiducial value from standard population synthesis models. More massive and more compact clusters have a lower critical mass-loading (smaller $\beta_\mathrm{crit}$ and $\beta_\mathrm{crit,min}$) for cooling, due to a greater number of massive stars depositing more wind material in a smaller volume. As an aside, note that $\beta_\mathrm{crit}$ in equation~(\ref{eq:beta_crit_scaled}) diverges as $\Delta R\rightarrow 0$ because winds from stars at the edge of the cluster always drive a cluster outflow, but with smaller and smaller $\dot M_\mathrm{esc}$ as $R_\mathrm{cool}\rightarrow R_{cl}$. Formally, no amount of mass-loading can make $R_\mathrm{cool}=R_{cl}$ in the limit of an infinite number of sources, provided constant $\dot E_w$ and $\dot M_w$ throughout $R_{cl}$. However, even $\beta\sim2\times\beta_\mathrm{crit,min}$ yields a cluster where a large fraction of the volume cools. Note also that $R_{cl}$ does not necessarily have to be the edge of the GC; it is the edge of the region where massive star winds are depositing mass and energy. If there is mass segregation for the massive stars, as seen in e.g. 47 Tuc \citep{Zhang2015}, $R_{cl}$ could be smaller than the size of the GC we see today, which would decrease $\beta_\mathrm{crit}$ and $\beta_\mathrm{crit,min}$.

The deposited wind mass that does not cool and escapes the cluster as a wind is $\dot M_\mathrm{esc}=\beta\dot M_w(1-R_\mathrm{cool}^3/R_{cl}^3)$. For $\beta\lesssim\beta_\mathrm{crit,min}$, $\dot M_\mathrm{esc}\sim \beta\dot M_w$. As $\beta$ increases, $R_\mathrm{cool}\rightarrow R_{cl}$, and $\dot M_\mathrm{esc}\ll\beta\dot M_w$. The cluster wind kinetic luminosity,
\begin{align}
&\dot E_\mathrm{esc}=\frac{1}{2}\dot M_\mathrm{esc} v_\infty^2 \nonumber \\
&\approx 10^{36}\ \mathrm{ergs\ s}^{-1}\ \frac{\alpha_{0.1}^{2.35}}{\beta_2^{1.85}} \nonumber \\
&\times\left(\frac{\dot E_w}{10^{37.8}\ \mathrm{ergs\ s}^{-1}}\right)^{2.35}\left(\frac{\dot M_w}{10^{-3.1}\ M_\odot\ \mathrm{yr}^{-1}}\right)^{-1.85}\left(\frac{R_{cl}}{1\ \mathrm{pc}}\right)^{0.5}, \label{eq:Edot_esc}
\end{align}
and momentum ejection rate,
\begin{align}
&\dot p_\mathrm{esc}=\dot M_\mathrm{esc}v_\infty \nonumber \\
&\approx 3\times10^{29}\ \mathrm{dynes\ s}^{-1}\ \frac{\alpha_{0.1}^{1.85}}{\beta_2^{1.35}} \nonumber \\
&\times\left(\frac{\dot E_w}{10^{37.8}\ \mathrm{ergs\ s}^{-1}}\right)^{1.85}\left(\frac{\dot M_w}{10^{-3.1}\ M_\odot\ \mathrm{yr}^{-1}}\right)^{-1.35}\left(\frac{R_{cl}}{1\ \mathrm{pc}}\right)^{0.5}, \label{eq:pdot_esc}
\end{align}
are thus dramatically reduced for $\beta\gtrsim\beta_\mathrm{crit,min}$, since $\dot M_\mathrm{esc}$ rapidly decreases as $R_\mathrm{cool}\rightarrow R_{cl}$. We thus hypothesize that for $\beta\gtrsim\beta_\mathrm{crit,min}$, the cluster is much more likely to retain ambient gas from the cluster formation process. Therefore, although highly dependent on $\alpha$ and $\beta$, equation~(\ref{eq:M1dRcrit}) gives the critical value of $M_1/R_{cl}$ above which second generation star formation may occur.

\begin{figure}
\includegraphics[width=\linewidth]{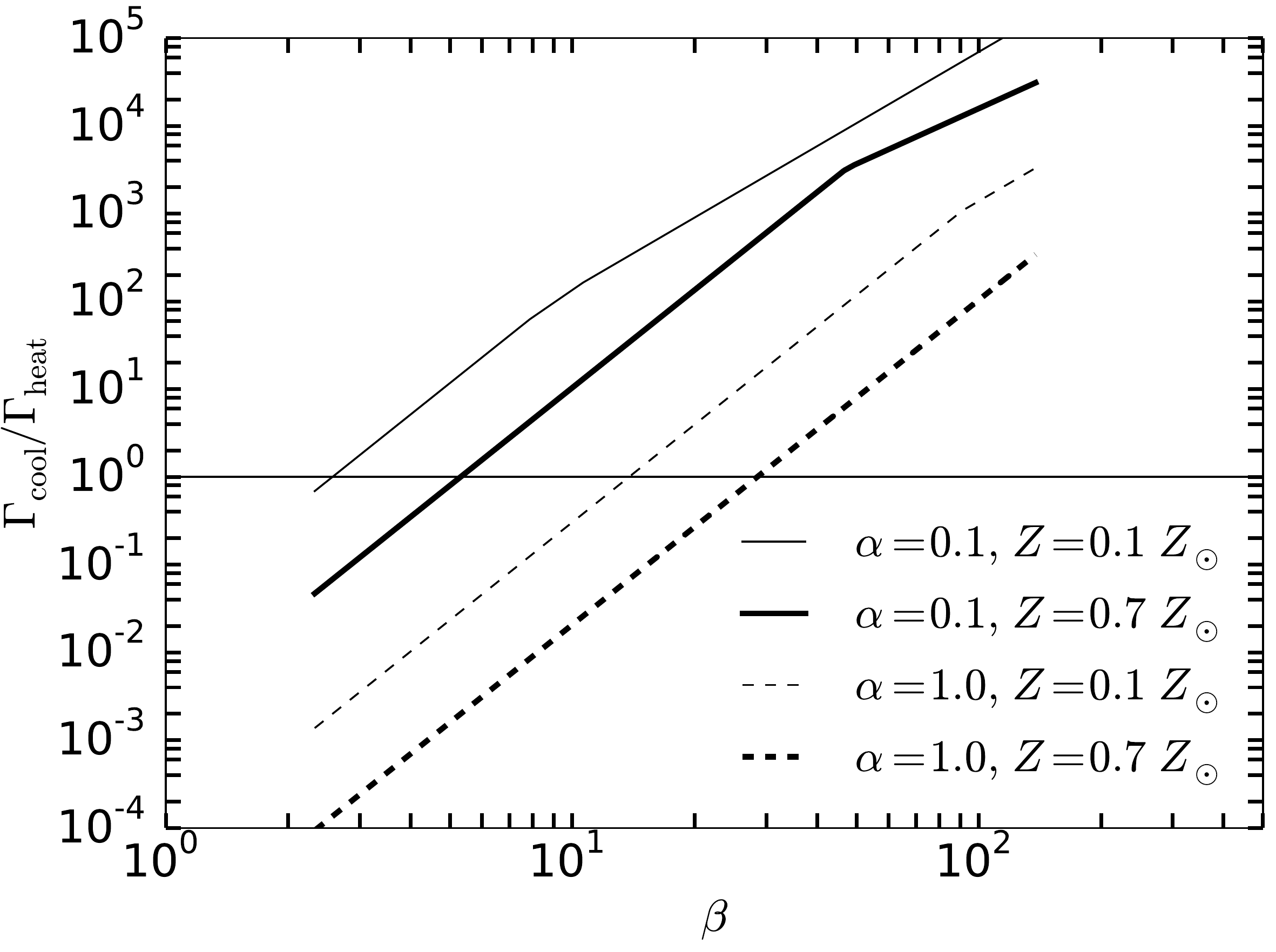}
\caption{The ratio of cooling rate ($\Gamma_\mathrm{cool}$) to heating rate ($\Gamma_\mathrm{heat}$) of massive star winds within 80\% of the cluster's volume ($R_\mathrm{cool}=0.93R_{cl}$) as a function of $\beta$, for energy thermalization efficiencies $\alpha=0.1$ (solid) and $\alpha=1.0$ (dashed) and metallicities one-tenth solar (thin lines) and $0.7$ solar (thick lines). This is our fiducial cluster with mass and radius $10^5\ M_\odot$ and $R_{cl}=1$ pc.}
\label{fig:rates_ratio}
\end{figure}

Figure~\ref{fig:rates_ratio} shows the ratio of the cooling rate to the heating rate as a function of $\beta$ for our fiducial GC for two energy thermalization efficiencies and two metallicities. The cooling rates were calculated at $0.93R_{cl}$, which contains 80\% of the cluster's volume. $\Gamma_\mathrm{cool}/\Gamma_\mathrm{heat}=1$ at $\beta=\beta_\mathrm{crit}$. A cluster with a low energy thermalization efficiency and a low metallicity requires a lower $\beta$ to obtain radiative cooling of 80\% of its volume than a cluster with high $\alpha$ and high $Z$.

$\beta_\mathrm{crit}$ and $\beta_\mathrm{crit,min}$ both depend on the normalization of the cooling function to the power of $\approx-0.27$, so naively a higher metallicity will reduce the critical mass-loading value. However, clusters with higher metallicity also have higher $\dot M_w$ by a factor of 1.4 and higher $\dot E_w$ by a factor of 7.4 in STARBURST99 (see discussion after equations~\ref{eq:Edot} and~\ref{eq:Mdot}), and the combined effect seen in Figure~\ref{fig:rates_ratio} shows that lower metallicity clusters will have lower $\beta_\mathrm{crit}$. Most GCs have metallicities down to [Fe/H]$\sim-2$, whereas our fiducial GC has a metallicity of [Fe/H]$\sim-1$, so values of $\beta_\mathrm{crit,min}$ and $\beta_\mathrm{crit}$ may be lower than what we report in equations~(\ref{eq:beta_crit_scaled}) and~(\ref{eq:beta_crit_min}) for GCs with similar masses and radii.

\begin{figure*}
\begin{minipage}{175mm}
\centering
\includegraphics[width=\linewidth]{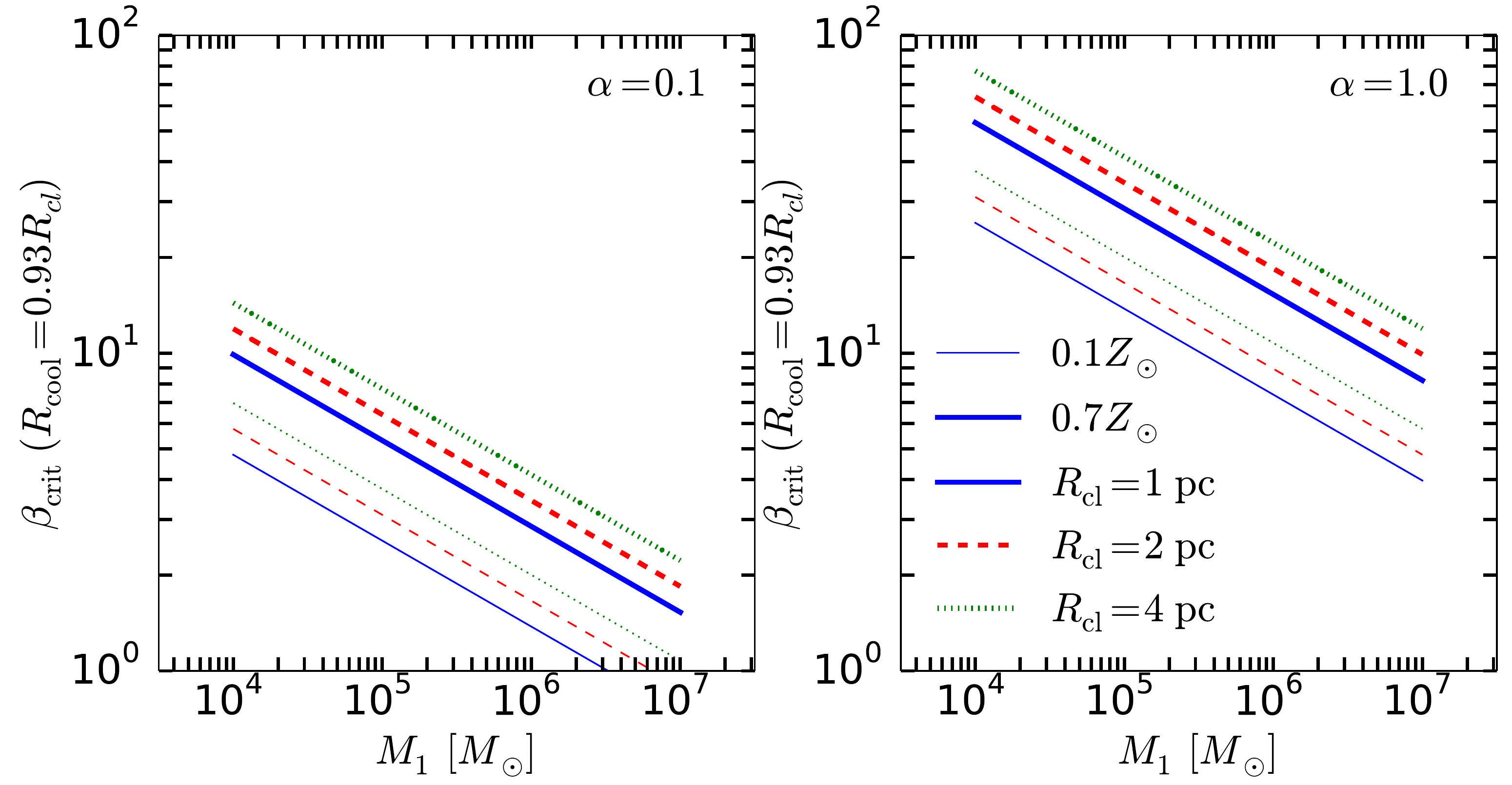}
\caption{The critical value of mass-loading $\beta$ required for the wind material deposited within 80\% of the cluster volume to cool radiatively as a function of first generation stellar mass $M_1$, for three cluster radii (blue solid for 1 pc, red dashed for 2 pc, green dotted for 4 pc), two cluster wind metallicities (heavy lines for $Z=0.7Z_\odot$, thin lines for $Z=0.1Z_\odot$), and two energy thermalization efficiencies ($\alpha=0.1$ in left panel, $\alpha=1$ in right panel).}
\label{fig:betacrit}
\end{minipage}
\end{figure*}

Figure~\ref{fig:betacrit} summarizes how the critical mass-loading depends on various properties of the cluster. Our fiducial cluster has $\beta_\mathrm{crit}\sim2$, so the temperature of the wind material and ambient gas mixture for $\beta=\beta_\mathrm{crit}$ is $T\approx 4\times10^5$ K, putting it firmly within the region of the cooling function dominated by metal-line cooling. Since more massive clusters have lower $\beta_\mathrm{crit}$, the temperature of wind material in these clusters is higher. A very massive and compact cluster with $M_1=10^7M_\odot$ and $R_{cl}=0.1$ pc has $\beta_\mathrm{crit}\approx0.3$ and thus $T\approx3\times10^6$ K, which is close to the minimum of the cooling function at $\sim10^7$ K, where this power-law approximation breaks down, and above which bremsstrahlung dominates. Thus, our model is more accurate for massive GCs, where a higher $\beta$ is necessary for cooling.

%%%%%%%%%%%%%%%%%%%%%%%%%%%%%%%%%%%%%%%%%%%%%%%%%%%%%%%%%%%%%%%%%%%%%%%%%%%%%%%%%%%%%%%%%
\subsection{Mass of the Second Generation}
\label{sec:2nd_mass}
%%%%%%%%%%%%%%%%%%%%%%%%%%%%%%%%%%%%%%%%%%%%%%%%%%%%%%%%%%%%%%%%%%%%%%%%%%%%%%%%%%%%%%%%%

As shown in equations~(\ref{eq:beta_crit_scaled}) and~(\ref{eq:beta_crit_min}), for $\beta=\beta_\mathrm{crit}$ or $\beta_\mathrm{crit,min}$, the mass deposition rate of massive stars must be approximately that of current models. However, a much larger $\beta$ is required for the second stellar generation to be as large as observed in globular clusters. If $\beta\approx \beta_\mathrm{crit}$ ($R_\mathrm{cool}\approx 0.93R_{cl}$) during the few Myr of post-main sequence massive star evolution, the total mass that cools to form a second generation would be $M_2\approx 10^{3.7}\ M_\odot$, implying a second generation mass fraction of $M_2/M_1\sim0.05$, too small to explain the large second generation masses observed. In \S\ref{sec:He_abundances}, we estimate a maximum upper bound on the wind mass loss rate and find that it can only be $\sim3$ times higher than equation~(\ref{eq:Mdot}), implying that winds alone can only produce a second generation $\sim0.05\times3\sim0.15$ as massive as the first generation\footnote{Our calculation agrees with the numerical results of \citet{Wunsch2017} (see their Figure 10) for $\alpha=0.1$ (their $\eta_\mathrm{he}=0.1$ and $\beta=1$ or 3 (their $\eta_\mathrm{ml}=0$ or 2).}.

To produce the large mass loadings required by the observations requires mixing with a substantial reservoir of ambient gas, presumably left over from first generation formation. We wish to give ourselves the flexibility to consider both enhanced mass loss rates during the post-main-sequence evolution of massive stars \emph{and} mixing with an amount of ambient gas in order to explore the relative importance of enrichment and dilution of He and other elements in \S\ref{sec:He_abundances} and \S\ref{sec:other_abundances}.

For these reasons, we write the second generation stellar mass as
\begin{equation}
M_2 = (M_\mathrm{gas}+\beta_\mathrm{wind}M_w)\left(\frac{R_\mathrm{cool}}{R_{cl}}\right)^3, \label{eq:M2_Mgas}
\end{equation}
where $\beta_\mathrm{wind}$ is the mass enhancement factor for the wind only, and $M_\mathrm{gas}$ is the mass of ambient gas mixed with the wind material within the cluster, and is given by
\begin{equation}
M_\mathrm{gas}=(\beta-\beta_\mathrm{wind})M_w, \label{eq:Mgas}
\end{equation}
where $M_w=\dot M_w \Delta t_\mathrm{SN}$ is the mass of wind material that has accumulated in the cluster over the time $\Delta t_\mathrm{SN}$, which is the time from the onset of strong mass deposition from winds at $2.5$ Myr after formation until the first core-collapse supernova that does \emph{not} collapse directly to a black hole occurs. Throughout the rest of this paper, we use $\Delta t_\mathrm{SN}=4.5$ Myr in our estimates, although the time until the first non-black-hole supernova is both uncertain \citep[e.g.][]{Pejcha2015,Sukhbold2016} and may be extended by rapidly rotating massive stars \citep{Ekstrom2012}. Although we scale to $\Delta t_\mathrm{SN}=4.5$ Myr throughout, in principle the second generation can form faster. Below we show that the cooling and freefall times are typically smaller than, but of order, $\Delta t_\mathrm{SN}\sim4.5$ Myr.

In this picture, $\beta_\mathrm{wind}=1$ corresponds to winds with $\dot M_w$ given by equation~(\ref{eq:Mdot}), and $\beta>\beta_\mathrm{wind}$ corresponds to mixing with ambient gas. In \S\ref{sec:He_abundances}, we consider $\beta_\mathrm{wind}$ as large as $\sim3$, with a range of ambient gas masses from $\beta=\beta_\mathrm{wind}$ (no mixing) to $\beta=30-100$, where the latter values of $\beta$ are needed to match the second generation stellar mass fractions observed in globular clusters. For a given value of $\beta_\mathrm{wind}$, equation~(\ref{eq:Mgas}) gives us a convenient way to represent the mass loading required to generate a given second generation stellar mass $M_2$. In particular, we can quote both a mass loading parameter $\beta$ and a corresponding mass of gas $M_\mathrm{gas}$ required to precipitate second generation formation of mass $M_2$. Until \S\ref{sec:He_abundances}, we consider only $\beta_\mathrm{wind}=1$.

\begin{figure}
\includegraphics[width=\linewidth]{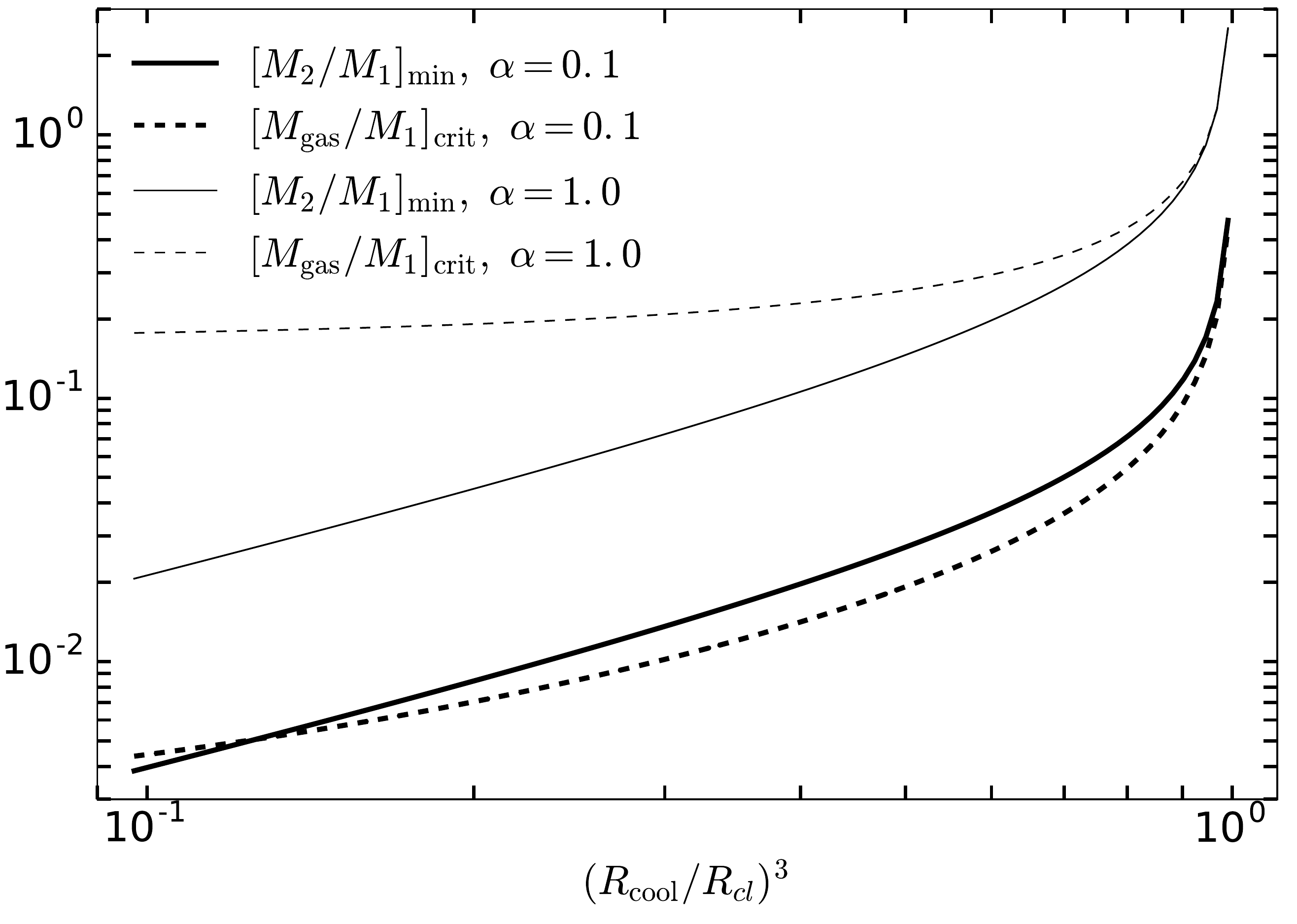}
\caption{The critical minimum mass of ambient gas necessary for cooling $[M_\mathrm{gas}/M_1]_\mathrm{crit}$ (dashed) when $\beta_\mathrm{wind}=1$ and the resultant minimum second stellar generation mass at the critical ambient gas mass $[M_2/M_1]_\mathrm{min}$ (solid) as a function of the fraction of the cluster's volume that radiatively cools, $(R_\mathrm{cool}/R_{cl})^3$. The model with efficiency of energy thermalization $\alpha=0.1$ is shown with thick lines and the model with $\alpha=1$ is shown with thin lines. If $M_\mathrm{gas}=M_\mathrm{gas,crit}$, then $M_2=M_\mathrm{2,min}$. If $M_\mathrm{gas}>M_\mathrm{gas,crit}$, then $M_2>M_\mathrm{2,min}$. The dashed curve corresponds to $\beta=\beta_\mathrm{crit}$ in equation~(\ref{eq:Mgas}), and the solid curve corresponds to $M_\mathrm{gas}=M_\mathrm{gas,crit}$ in equation~(\ref{eq:M2_Mgas}).}
\label{fig:M2_Mgas_vs_Vol}
\end{figure}

If $\beta<\beta_\mathrm{crit,min}$, then $M_2=0$ because none of the gas radiatively cools, and thus it all escapes the cluster. Figure~\ref{fig:M2_Mgas_vs_Vol} shows the critical $M_\mathrm{gas}/M_1$ ratio required for cooling when $\beta_\mathrm{wind}=1$ at each fraction of the total cluster volume as the dashed curves for $\alpha=0.1$ (thick) and $\alpha=1$ (thin). The second generation mass that forms when $M_\mathrm{gas}/M_1$ is equal to the critical value for cooling is the \emph{minimum} value of $M_2/M_1$ (solid curves, thick for $\alpha=0.1$ and thin for $\alpha=1$). If $\beta>\beta_\mathrm{crit,min}$ or, alternatively, if $M_\mathrm{gas}>M_\mathrm{gas,crit}$, the more massive the second generation will be. In order to have $M_2/M_1\sim$ few, we require $\beta\gg\beta_\mathrm{crit,min}$ (thus $M_\mathrm{gas}/M_1\gg[M_\mathrm{gas}/M_1]_\mathrm{crit}$). For $(R_\mathrm{cool}/R_{cl})^3\sim0.8$ and $M_\mathrm{gas}/M_1\sim3$, $\beta\sim100$. The $\alpha=1$ $[M_\mathrm{gas}/M_1]_\mathrm{crit}$ curve falls above the $\alpha=1$ $[M_2/M_1]_\mathrm{crit}$ curve because more mass loading is necessary for the gas to cool when the energy thermalization efficiency is high, but not all of this gas enters into the second generation. Only a fraction $(R_\mathrm{cool}/R_{cl})^3$ of the wind and ambient gas mixture forms the second generation.

For $\beta\sim100$, the temperature of the thermalized gas is decreased by a factor of $100$ (equation~\ref{eq:temp_scaled}), which makes it $T\sim8\times10^3$ K for $\alpha=0.1$ and $T\sim8\times10^4$ K for $\alpha=1$. For $\alpha=0.1$, the temperature is low enough that radiatively cooling will not lower the temperature further than the $\sim10^4$ K photoionization maintains, so cooling is not necessary for winds to be retained in the cluster. For $\alpha=1$, the temperature is in the low-temperature regime of the cooling function that scales as $0.3$ with temperature for one-tenth solar metallicity, so equation~(\ref{eq:beta_crit_scaled}) does not apply. However, our result that the thermalized gas radiatively cools is unchanged: a derivation of $\beta_\mathrm{crit}$ for this region of the cooling function gives $\beta_\mathrm{crit}\sim1$ for $T<10^5$ K. $\beta\sim100$ is far larger than this $\beta_\mathrm{crit}$, so the thermalized gas in both cases of $\alpha=0.1$ and $\alpha=1$ will be cool enough to be retained within the cluster.

Radiative cooling is necessary, but not sufficient for the gas to form a second generation of stars. The massive stars in globular clusters output high fluxes of ionizing photons that, if the gas is not shielded, will maintain a temperature of $\sim10^4$ K even in the cooled gas. Star formation requires cooler temperatures, so the radiatively cooled gas cannot be continuously ionized by the starlight in the cluster if a second stellar generation is to form \citep{Palous2014}. \citet{Wunsch2017} show that the amount of mass that cools is consistently greater than the amount of mass necessary to shield a central clump of gas from the ionizing photons for $\alpha=0.1$ ($\alpha=1$ was not examined), thus allowing it cool below $10^4$ K and form stars. We assume in our calculation that the amount of gas that cools is equal to the amount of gas that enters into the second generation. A model with smaller second generation star formation efficiencies will require proportionally more gas to produce a second stellar generation of a given size. Future numerical works should further assess this assumption and the ability of gas to self-shield.

\begin{figure*}
\begin{minipage}{175mm}
\centering
\includegraphics[width=\linewidth]{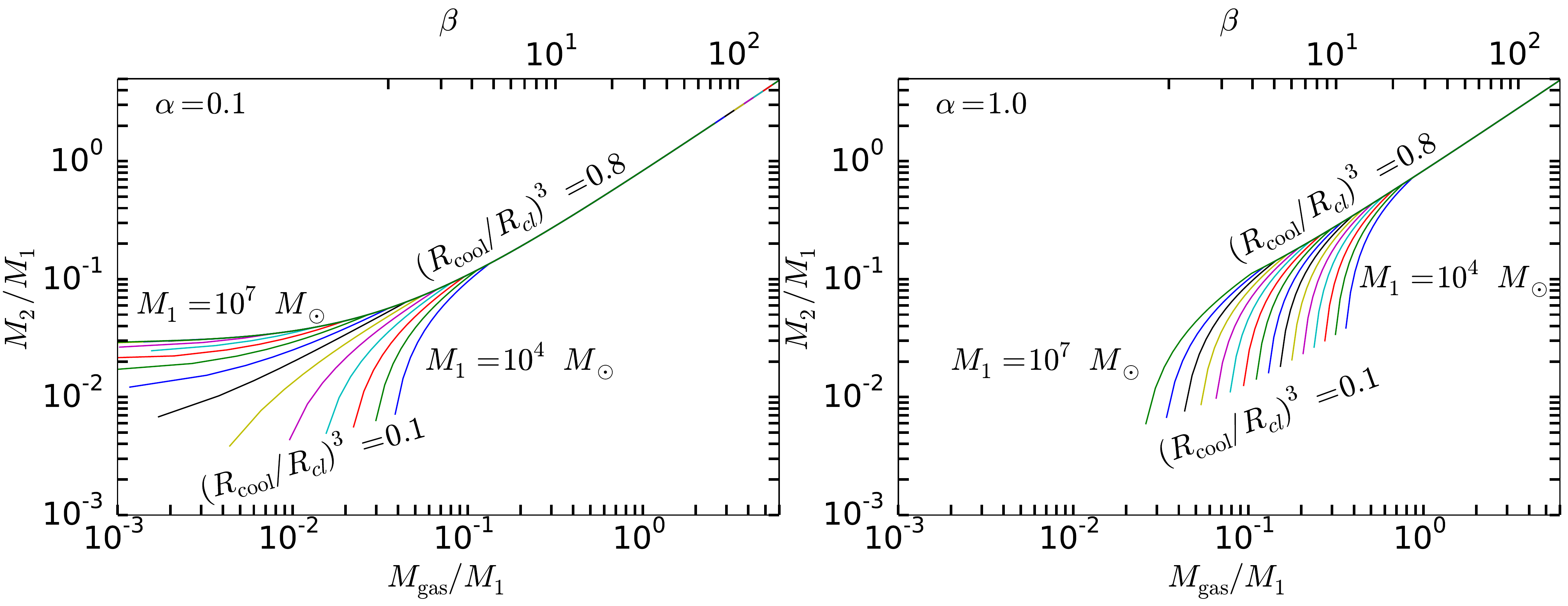}
\caption{The ratio of the second generation's stellar mass $M_2$ to the first generation's stellar mass $M_1$ as a function of the ratio of ambient gas mass $M_\mathrm{gas}$ to the first generation stellar mass when $\beta_\mathrm{wind}=1$. Curves show different first generation masses, from $10^4\ M_\odot$ to $10^7\ M_\odot$, in multiples of $10^{0.2}$. Left panel shows models with energy thermalization efficiency $\alpha=0.1$ and right panel shows models with $\alpha=1$. All GCs have radius $R_{cl}=1$ pc, one-tenth solar metallicity, and have been accumulating wind material for $\Delta t_\mathrm{SN}=4.5$ Myr. Top axis shows $\beta$ values corresponding to $M_\mathrm{gas}/M_1$ values on bottom axis.}
\label{fig:M2dM1}
\end{minipage}
\end{figure*}

Figure~\ref{fig:M2dM1} shows the ratio of the stellar mass in the second generation to that in the first generation $M_2/M_1$ as a function of the ratio of ambient gas mass to first-generation stellar mass within the cluster when $\beta_\mathrm{wind}=1$, for several cluster models with different first generation stellar masses. The upper edge limit of each curve corresponds to 80\% of the cluster's volume cooling ($R_\mathrm{cool}=0.93R_{cl}$ in equation~\ref{eq:M2_Mgas}), where $M_2/M_1\propto M_\mathrm{gas}/M_1$. The lower edge limit, where the curves end, corresponds to the inner 10\% of the cluster volume ($R_\mathrm{cool}=0.46R_\mathrm{cl}$). We avoid edge and center effects by focusing on $0.1-0.8$ of the cluster's volume (see discussion of equations~\ref{eq:beta_crit},~\ref{eq:beta_crit_scaled}, and~\ref{eq:beta_crit_min}). Cooling and second generation star formation set in at lower $M_\mathrm{gas}/M_1$ for higher $M_1$ because these GCs have a higher number density of massive stars depositing winds, and therefore do not require as much mass loading to reach the high densities necessary for radiative cooling. We explore other values of $R_{cl}$, but find our large-$\beta$ results remain unchanged for 1 pc $<R_{cl}<$ 10 pc, so we focus on $R_{cl}=1$ pc for the remainder of this paper.

\begin{figure*}
\begin{minipage}{175mm}
\centering
\includegraphics[width=\linewidth]{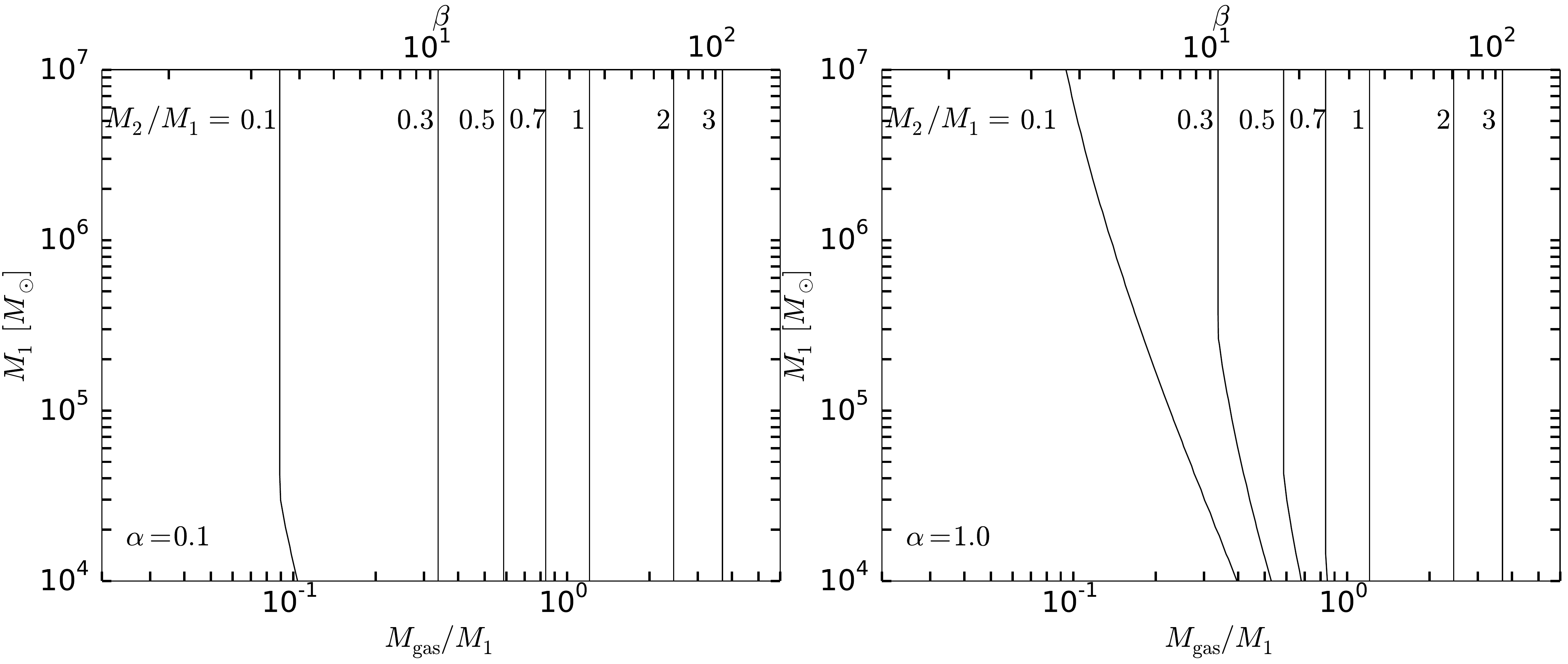}
\caption{Contour plot of the ratio of stellar mass in the second generation to the first generation, $M_2/M_1$, as a function of both the first generation stellar mass $M_1$ and ratio of gas mass to first generation stellar mass, $M_\mathrm{gas}/M_1$, when $\beta_\mathrm{wind}=1$. Models in left panel have energy thermalization efficiency $\alpha=0.1$ and models in right panel have $\alpha=1$. All clusters shown here have radius $R_{cl}=1$ pc, one-tenth solar metallicity, and have been accumulating wind material for $t_\mathrm{SN}=4.5$ Myr. Top axis shows $\beta$ values corresponding to $M_\mathrm{gas}/M_1$ values on bottom axis.}
\label{fig:contour}
\end{minipage}
\end{figure*}

The contours in Figure~\ref{fig:contour} show the ratio $M_2/M_1$, as a function of both first generation stellar mass and ratio of ambient gas mass to first-generation stellar mass when $\beta_\mathrm{wind}=1$, for the same cluster models as in Figure~\ref{fig:M2dM1}. In order to produce a second stellar generation $\sim0.5-3$ times the size of the first, a mass of ambient gas of order the mass of the first stellar generation or larger is required, regardless of the mass of the first generation. There is very little dependence of $M_2/M_1$ on the first-generation stellar mass of the cluster, as implied by all lines becoming vertical at large $M_2/M_1$ in Figure~\ref{fig:contour}. Only $M_\mathrm{gas}/M_1$ has much effect on $M_2/M_1$ in this limit. Certain combinations of $M_1$ and $M_\mathrm{gas}$ simply do not produce a second generation of stars at all, because the stellar winds combined with ambient gas never reach high enough density to raise the cooling rate above the heating rate ($\beta\lesssim\beta_\mathrm{crit,min}$). This parameter space is the far left of the plot in the left panel, and pushes further to the right in the bottom of the right panel, where $\alpha=1$. This is another visualization of the effect seen in Figure~\ref{fig:M2dM1}, where $\beta_\mathrm{crit,min}\sim5$ for $\alpha=1$ implies that some amount of mass loading is always necessary for cooling to occur when the energy thermalization efficiency is high.

\begin{figure*}
\begin{minipage}{175mm}
\centering
\includegraphics[width=\linewidth]{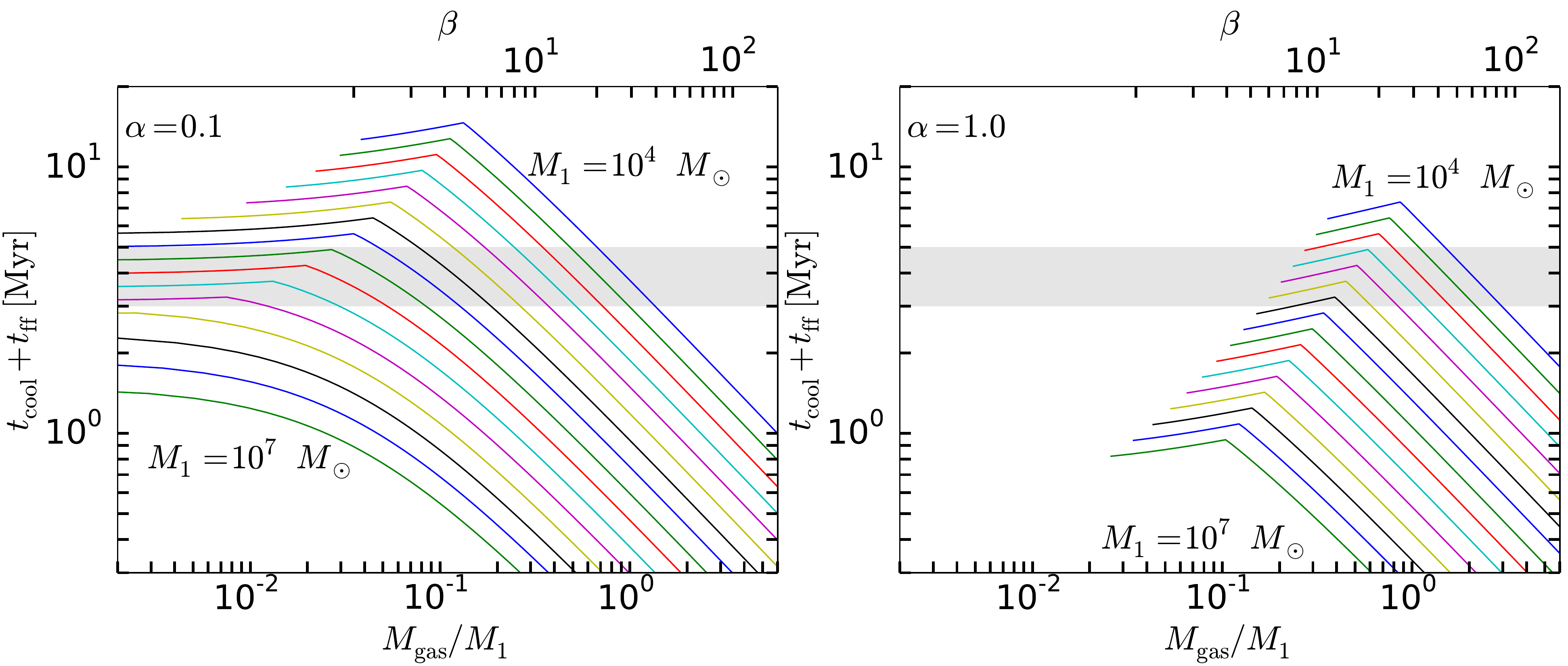}
\caption{The sum of the cooling time and the free fall time for GCs with first generations in the mass range $10^4\ M_\odot$ to $10^7\ M_\odot$, in multiples of $10^{0.2}$, as a function of the ratio of ambient gas to first generation stellar mass when $\beta_\mathrm{wind}=1$. Left panel shows models with energy thermalization efficiency $\alpha=0.1$ and right panel shows models with $\alpha=1$. These are all clusters with $R_{cl}=1$ pc, $Z=0.1Z_\odot$, and they have been accumulating material from stellar winds for $\Delta t_\mathrm{SN}=4.5$ Myr. Horizontal gray band at $3-5$ Myr shows the approximate time after winds turn on at which the most massive stars end their lives \emph{without} direct collapse to black hole. Top axis shows values of $\beta$ that correspond to values of $M_\mathrm{gas}/M_1$ on bottom axis.}
\label{fig:tff}
\end{minipage}
\end{figure*}

Figure~\ref{fig:tff} shows the formation time of the second generation as the sum of the cooling time and the free-fall time (calculated as $t_\mathrm{ff}=\sqrt{3\pi/(32G\mu m_p n)}$ with density $n$ given by equation~\ref{eq:dens}) as a function of $M_\mathrm{gas}/M_1$ for the same cluster models from Figures~\ref{fig:M2dM1} and~\ref{fig:contour}\footnote{We calculate $t_\mathrm{ff}$ assuming constant density within $R_\mathrm{cool}$, but the shock-compressed cooling gas can have much higher density and lower free-fall time. This would allow the second generation to form faster than we predict, and thus our estimates are upper limits on the time to form a second generation.}. We assume the first generation forms instantaneously, but that significant post-main sequence massive star winds do not turn on until $2.5$ Myr. The cooling time is given by
\begin{equation}
t_\mathrm{cool}=\frac{3}{2}\frac{P}{\Gamma_\mathrm{cool}}
\end{equation}
where $P$ is the pressure of the thermalized wind ejecta and ambient gas mixture.

The horizontal gray band at $3-5$ Myr in Figure~\ref{fig:tff} shows the approximate time after the onset of winds at $2.5$ Myr when the first supernova from the most massive star in the first generation can be expected to occur $3-5$ Myr after star formation, when the winds turn on. For the more massive clusters, $t_\mathrm{cool}+t_\mathrm{ff}<3$ Myr for all $M_\mathrm{gas}/M_1$, justifying that star formation can occur rapidly, before supernovae. The ``kink" in the curves is where $(R_\mathrm{cool}/R_{cl})^3=0.8$, where the cooling radius is then held constant instead of increasing further toward $R_{cl}$. Any values of $M_\mathrm{gas}/M_1$ greater than this point for a given curve have $M_\mathrm{gas}/M_1>[M_\mathrm{gas}/M_1]_\mathrm{crit}$, and we see that larger values of $M_\mathrm{gas}/M_1$ reduce $t_\mathrm{cool}+t_\mathrm{ff}$ below 3 Myr for all clusters. Note that parameter regimes with $t_\mathrm{ff}+t_\mathrm{cool}>\Delta t_\mathrm{SN}$ would either occur in low mass clusters or when $M_\mathrm{gas}/M_1$ is small enough that it will not produce a sizable second generation. For $\alpha=1$, very few combinations of $M_1$ and $M_\mathrm{gas}/M_1$ do not cool and collapse before supernovae begin because $M_\mathrm{gas}/M_1\sim1$ is necessary for cooling to occur at all, so if cooling does occur, there is a high gas density and the free fall and cooling times are short. For our fiducial GC with $M_1=10^5\ M_\odot$, $R_{cl}=1$ pc, $\alpha=0.1$, and one-tenth solar metallicity, Figure~\ref{fig:contour} shows $M_\mathrm{gas}/M_1\sim0.5-4$ is necessary to form a second generation $0.5-3$ times as massive as the first when $\beta_\mathrm{wind}=1$. This means our fiducial cluster would form a second generation in $t_\mathrm{cool}+t_\mathrm{ff}\approx 1-3$ Myr --- before core collapse supernovae begin.

%%%%%%%%%%%%%%%%%%%%%%%%%%%%%%%%%%%%%%%%%%%%%%%%%%%%%%%%%%%%%%%%%%%%%%%%%%%%%%%%%%%%%%%%%
\subsection{Comparison with Observations}
\label{sec:obs}
%%%%%%%%%%%%%%%%%%%%%%%%%%%%%%%%%%%%%%%%%%%%%%%%%%%%%%%%%%%%%%%%%%%%%%%%%%%%%%%%%%%%%%%%%

\begin{figure}
\includegraphics[width=\linewidth]{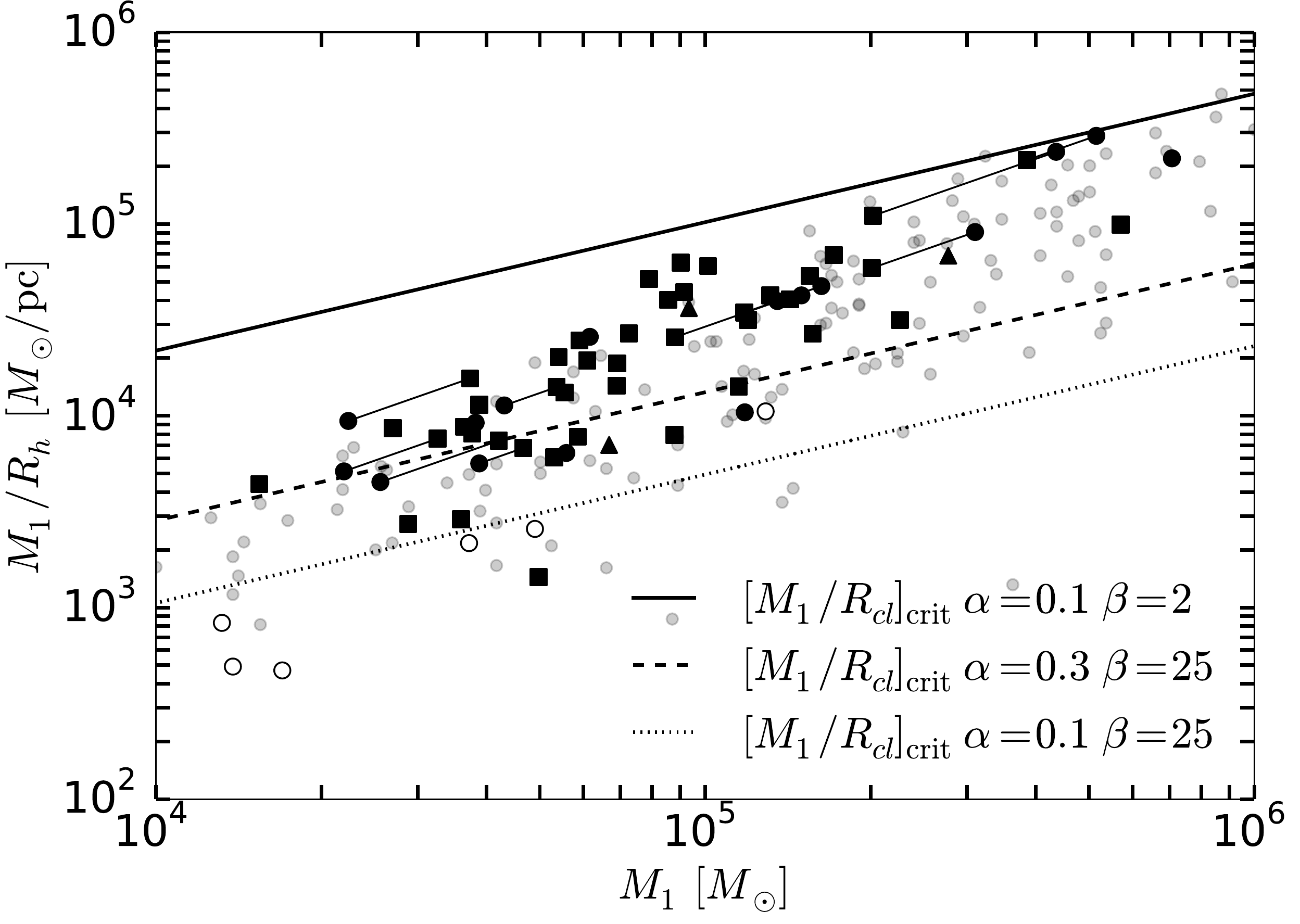}
\caption{The ratio of mass in the first stellar generation $M_1$ to the cluster's half-light radius $R_h$ as a function of $M_1$. Solid, dark points indicate clusters where a second stellar generation has been identified: circles are clusters from \citet{Carretta2009a}, squares are clusters from \citet{Milone2017}, and triangles are clusters from \citet{Niederhofer2016a,Niederhofer2016b}. Open circles are clusters that are identified as candidates for only first generation or mostly first generation by \citet{Caloi2011}. Light gray solid circles are all other clusters in the \citet{McLaughlin2005} database not included in the aforementioned six studies, for which we assume the first generation stellar mass is equal to the total stellar mass of the cluster for the purposes of plotting. Lines connecting solid dark points indicate that the multiple points represent a single cluster for which different studies found different first generation fractions. Solid, dashed, and dotted lines show the minimum value of $M_1/R_{cl}$ a cluster must have for nearly all of it to cool and form a second generation, for three combinations of $\alpha$ and $\beta$ (see equation~\ref{eq:M1dRcritfullcool}), where we assume $R_{cl}=R_h$ (see caveats in text).}
\label{fig:MdRvsM}
\end{figure}

With an understanding of how the various GC parameters affect the mass of the second generation of stars, we can now compare our model to observations of multiple stellar generations in GCs. Several studies \citep[e.g.][]{Carretta2011,Carretta2014,Gratton2015,Milone2017} have examined the spread of light element abundances in Galactic GCs to determine which stars are members of first and second stellar generations. The ratio of second generation to first generation stars can be as low as $M_2/M_1\sim0.35$ \citep{Boberg2016} or $M_2/M_1\sim0.5$ \citep{Milone2017}, but are more typically $M_2/M_1\sim2-3$ \citep{Carretta2009a}. Using the GC masses and half-light radii from the database presented in \citet{McLaughlin2005} and the first generation fractions from \citet{Carretta2009a,Niederhofer2016a,Niederhofer2016b,Milone2017}, we plot $M_1/R_h$ as a function of $M_1$ in Figure~\ref{fig:MdRvsM} as the dark, solid points, where $M_1$ is the stellar mass of the first generation and $R_h$ is the half-light radius. We assume that the initial mass functions of both generations are equivalent and well-sampled so that the fraction of stars in each generation is equivalent to the fraction of initial stellar mass in each generation. The open points are clusters that were identified as being only-first generation or mostly-first generation by \citet{Caloi2011}, but we plot only those clusters in their sample that were \emph{not} later found to be hosting multiple populations: NGC 6235, AM 1, ARP 2, PAL 3, PAL 4, and PAL 14. The light gray solid points are all other clusters in the \citet{McLaughlin2005} database. Lines connecting dark solid points indicate that multiple studies examined a particular cluster and determined different values for the first generation fraction, thus giving different values of $M_1$ for the same cluster. With the exception of a few outliers, it seems that those clusters that are made of only- or mostly-first generation stars have lower values of $M_1/R_h$ at all $M_1$ than those clusters with an identified second generation.

Figure~\ref{fig:MdRvsM} also shows the critical condition on $M_1/R_{cl}$ for nearly all of the cluster's volume to cool (see equation~\ref{eq:M1dRcritfullcool}), for a few different values of $\alpha$ and $\beta$, as the solid, dashed, and dotted lines. If a cluster has the specified values of $\alpha$ and $\beta$, then our model predicts it would form a second generation if it falls above the line, and would not if it falls below the line. The location of the model lines in this plot are highly sensitive to the choice of $\alpha$ and $\beta$, which are not observationally well-constrained. The dotted line, which shows $[M_1/R_{cl}]_\mathrm{crit}$ for $\alpha=0.1$ and $\beta=25$, splits the clusters into those with measured second generation fractions and those that are only or mostly first generation. Our model predicts that a cluster with these parameters would produce a second generation of mass $M_2\sim0.7M_1$ if its $M_1/R_{cl}$ falls above this line.

Comparing the model lines to the data in Figure~\ref{fig:MdRvsM} has several caveats. First, we have assumed that $R_{cl}$, the radius within which the first generation's massive stars deposit their winds, is equal to the observed half-light radius $R_h$, but the first generation massive stars could have been centrally-concentrated. Second, the fraction of second generation stars is not well-constrained, as different methods for separating the two generations yield sometimes different fractions, and this changes where a cluster falls in this figure because it directly affects $M_1$. For example, NGC 6809 was found to have a first generation fraction of $M_1/(M_1+M_2)\approx0.20\pm0.05$ by examining the sodium-oxygen anticorrelation by \citet{Carretta2009a}, and $M_1/(M_1+M_2)\approx0.311\pm0.029$ by examining the split red giant branch in a two-color map by \citet{Milone2017}. Third, our model's predictions are strongly dependent on the values of $\alpha$ and $\beta$, which may vary from cluster to cluster with ambient gas mass, metallicity, or cluster environment.

\begin{figure*}
\begin{minipage}{175mm}
\includegraphics[width=\linewidth]{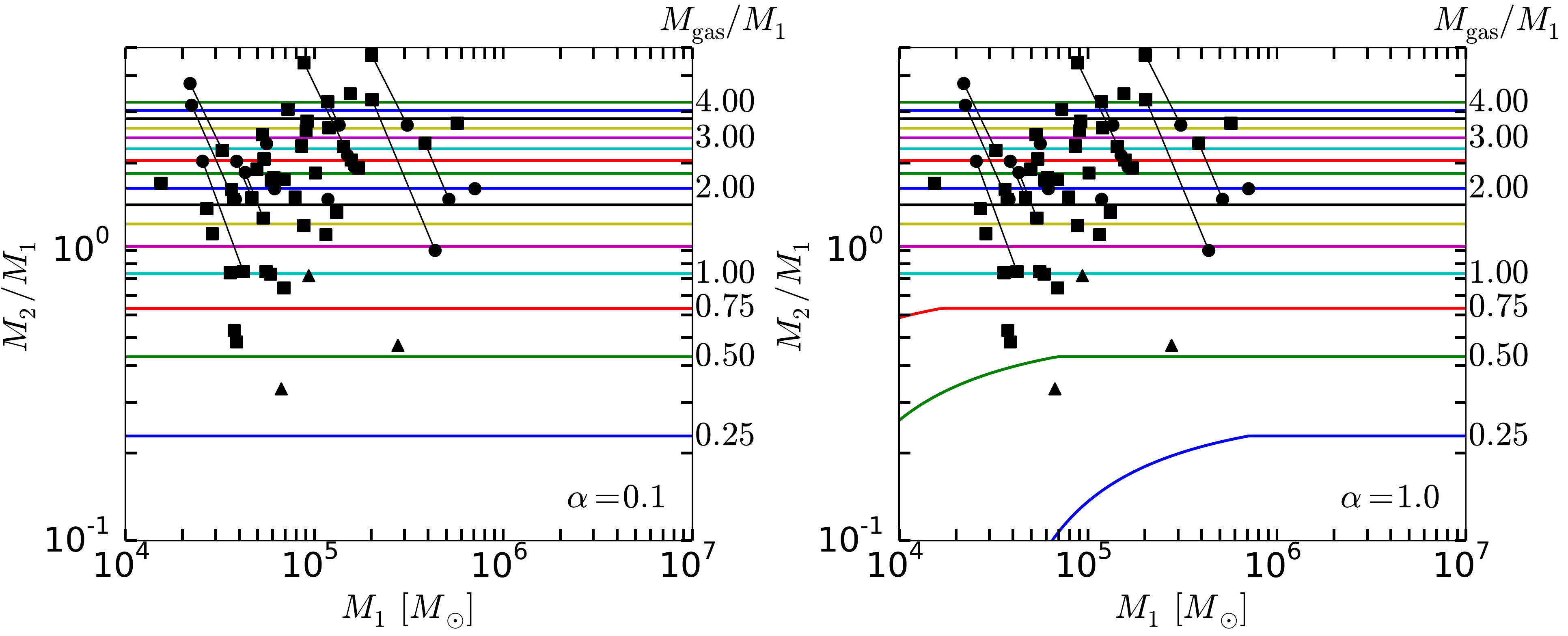}
\caption{Colored lines indicate the ratio of second generation stellar mass to first generation stellar mass in our fiducial model as a function of first generation stellar mass. Left panel shows models with energy thermalization efficiency $\alpha=0.1$, and right panel shows models with $\alpha=1$. $M_2/M_1$ increases from $M_2/M_1\sim0.2$ to $M_2/M_1\sim3$ as $M_\mathrm{gas}/M_1$ increases from $0.25$ to $4$ when $\beta_\mathrm{wind}=1$, as indicated by labels for select curves to right of plot. Filled circles are clusters examined by \citet{Carretta2009a}, filled squares are those examined by \citet{Milone2017}, and filled triangles are those examined by \citet{Niederhofer2016a,Niederhofer2016b}. Lines connecting points indicate where a cluster was found by different studies to have different first generation fractions, thus changing both $M_1$ and $M_2/M_1$ for that cluster. Curves were produced by our model with one-tenth solar metallicity and have accumulation of wind material for $\Delta t_\mathrm{SN} = 4.5$ Myr.}
\label{fig:M2dM1data}
\end{minipage}
\end{figure*}

Importantly, most clusters where the second generation has been studied are high-mass. Figure~\ref{fig:M2dM1data} and the critical condition of equation~(\ref{eq:M1dRcritfullcool}) imply that a cluster with low $M_1/R_{cl}$ cannot form a second generation because the winds do not meet the critical condition for radiative cooling and retention of the wind material. Verifying any mass dependence of the presence of a second generation requires studying low-mass clusters as well.

Figure~\ref{fig:M2dM1data} shows the ratio of mass in the second stellar generation to that in the first for the same clusters indicated by the dark solid points in Figure~\ref{fig:MdRvsM} and values of $M_2/M_1$ for different values of $M_\mathrm{gas}/M_1$ when $\beta_\mathrm{wind}=1$ (right vertical axis), as functions of $M_1$. There is no variation in our model with $M_1$ for the values of $M_\mathrm{gas}/M_1$ shown (see Figure~\ref{fig:contour}) for $\alpha=0.1$, and only some variation for the lower values of $M_2/M_1$ for $\alpha=1$, and there is also little variation in $M_2/M_1$ with $M_1$ in the data. Points and connecting lines have the same meaning as in Figure~\ref{fig:M2dM1data}.

In order to produce a second generation $0.5-3$ times as large as the first, $M_\mathrm{gas}/M_1=0.5-4$ is necessary if $\beta_\mathrm{wind}=1$. For this range of $M_\mathrm{gas}/M_1$ of our fiducial cluster ($M_1=10^5\ M_\odot$), the final mass after the second generation has formed is $M_1+M_2=1.5-5.5\times10^5\ M_\odot$. If post-main sequence wind mass loss rates are enhanced ($\beta_\mathrm{wind}>1$), the ambient gas mass required to form a second generation comparable in mass to the first is less than the ambient gas mass required when $\beta_\mathrm{wind}=1$.

%%%%%%%%%%%%%%%%%%%%%%%%%%%%%%%%%%%%%%%%%%%%%%%%%%%%%%%%%%%%%%%%%%%%%%%%%%%%%%%%%%%%%%%%%
\subsection{Helium Abundances}
\label{sec:He_abundances}
%%%%%%%%%%%%%%%%%%%%%%%%%%%%%%%%%%%%%%%%%%%%%%%%%%%%%%%%%%%%%%%%%%%%%%%%%%%%%%%%%%%%%%%%%

We now turn to the light-element abundance spreads of GCs, focusing on the helium enrichment of the second generation. Since our model thus far does not include abundance information beyond an overall metallicity, we adopt a model for wind ejecta from massive stars. We use the pre-supernova wind abundances for non-rotating solar metallicity stars from the recent core-collapse simulations of \citet{Sukhbold2016}. Because GC [Fe/H] values are $-2$ to $-1$ we assume the metallicity of the stars does not have a very strong effect on the relative helium content of the winds for the purposes of our estimate.

As winds from the most massive stars dominate the mass and energy deposition at early times in GCs, we focus on stars with masses $25-120\ M_\odot$. \citet{Sukhbold2016} calculate wind abundances for many stellar masses in the range $25-120\ M_\odot$, within which we linearly interpolate to obtain a finer stellar mass sampling. We integrate wind masses with a Salpeter IMF to obtain the total mass in winds produced by stars in our selected mass range. The IMF-integrated wind mass from \citet{Sukhbold2016} is not equal to the mass in winds, $M_w$, from STARBURST99, so we multiply the \citet{Sukhbold2016} yields by a correction factor of $\sim0.75$.

We calculate $\Delta Y=\Delta\frac{M_\mathrm{He}}{M_\mathrm{tot}}$ as
\begin{equation}
\Delta Y = \frac{M_\mathrm{He,wind}+M_\mathrm{He,gas}}{M_\mathrm{wind}+M_\mathrm{gas}} - Y_1,
\end{equation}
where $M_\mathrm{He,wind}$ is the mass of helium in the wind, $M_\mathrm{He,gas}$ is the mass of helium in the ambient gas, $M_\mathrm{wind}$ is the total mass of wind, and $Y_1$ is the $Y$ value of the first stellar generation. The first stellar generation has
\begin{equation}
Y_1=\frac{M_{\mathrm{He},\odot}}{M_{\mathrm{H},\odot} + M_{\mathrm{He},\odot} + Z_\mathrm{gas}\times M_{\mathrm{metals},\odot}},
\end{equation}
where $(M_{\mathrm{He},\odot}/M_{\mathrm{H},\odot})\simeq0.382$ and $(M_{\mathrm{metals},\odot}/M_{\mathrm{H},\odot})\simeq0.02$. For a gas metallicity of one-tenth solar, $Y_1\simeq0.276$. Since the ambient gas has the same elemental make-up as the first generation, this means $Y_\mathrm{gas}=M_\mathrm{He,gas}/M_\mathrm{gas}\simeq0.276$ as well. We find $M_\mathrm{He,wind}$ and $M_\mathrm{wind}$ from the yields produced by \citet{Sukhbold2016}.

Using equation~(\ref{eq:Mgas}), if $\beta_\mathrm{wind}=1$, then
\begin{align}
\Delta Y &= \frac{M_\mathrm{He,wind}+Y_\mathrm{gas}\dot M_w \Delta t_\mathrm{SN}(\beta-1)}{\dot M_w \Delta t_\mathrm{SN}+\dot M_w \Delta t_\mathrm{SN} (\beta-1)}-Y_1 \nonumber \\
&= \frac{Y_\mathrm{wind}+Y_\mathrm{gas}(\beta-1)}{\beta}-Y_1 \label{eq:delta_Y}
\end{align}
where $Y_\mathrm{wind}=M_\mathrm{He,wind}/M_\mathrm{wind}$. All dependence on first generation stellar mass drops out because $M_\mathrm{He,wind}\propto M_1$ and $M_\mathrm{wind}\propto M_1$. However, there is still dependence of $\Delta Y$ on the total mass of the cluster, $M_1+M_2$, because a higher $\beta$ indicates a higher $M_\mathrm{gas}$, which increases $M_2$ (see eq.~\ref{eq:M2_Mgas}). If $\beta_\mathrm{wind}>1$, we use equation~(\ref{eq:Mgas}) to write $\beta$ in terms of $\beta_\mathrm{wind}$, $\dot M_w\Delta t_\mathrm{SN}$, and $M_\mathrm{gas}$, and find
\begin{align}
&\Delta Y = \frac{(Y_\mathrm{wind}-Y_\mathrm{gas})(R_\mathrm{cool}/R_{cl})^3\beta_\mathrm{wind}\dot M_w \Delta t_\mathrm{SN} M_1^{-1}}{M_2/M_1}-Y_1 \nonumber \\
&\approx0.015\ \beta_\mathrm{wind,3}\left(\frac{M_1}{M_2}\right)\left(\frac{\dot M_w}{10^{-3.1}\ M_\odot\ \mathrm{yr}^{-1}}\right) \nonumber \\
&\times\left(\frac{R_\mathrm{cool}/R_{cl}}{0.93}\right)^3\left(\frac{\Delta t_\mathrm{SN}}{4.5\ \mathrm{Myr}}\right)\left(\frac{M_1}{10^5\ M_\odot}\right)^{-1} \label{eq:delta_Y_scaled}
\end{align}
where we have assumed $Y_\mathrm{gas}=Y_1=0.276$ and $Y_\mathrm{wind}=0.44$, and scaled to $\beta_\mathrm{wind}=3$, the maximum mass enhancement of the wind (see below). Equation~(\ref{eq:delta_Y_scaled}) only holds for $M_2\geq \beta_\mathrm{wind} M_\mathrm{wind}$, and $\Delta Y$ is maximum when $M_2=\beta_\mathrm{wind}M_\mathrm{wind}$ (see below).

Figure~\ref{fig:He} shows $\Delta Y$ in equation~(\ref{eq:delta_Y_scaled}) as a function of the ratio of second generation to first generation mass, for our fiducial massive star winds $\dot M_w$ when $\beta_\mathrm{wind}=1$ as the solid line. Producing a larger value of $M_2/M_1$ requires a larger amount of ambient gas, which dilutes the wind material and yields lower values of $\Delta Y$. The helium enhancement is $\Delta Y\sim0.002-0.013$ for $M_2/M_1\sim0.5-3$, which is produced by mixing with $M_\mathrm{gas}/M_1\sim0.5-4$.

Milky Way GCs have inferred helium enhancements of $\Delta Y\sim0.01-0.04$, with some as high as $\Delta Y\sim0.08$ or as low as $\Delta Y\sim0.001$, but these values are uncertain \citep{MacLean2016,Valcarce2014}. \citet{Bastian2015a} show no theoretical models can self-consistently produce all aspects of elemental abundance differences because there is a wide spread of helium enhancement among GCs, and we find our model cannot produce both inferred $\Delta Y$ values and measured $M_2/M_1$ values for GCs (shown as black points in Figure~\ref{fig:He}).

\begin{figure}
\includegraphics[width=\linewidth]{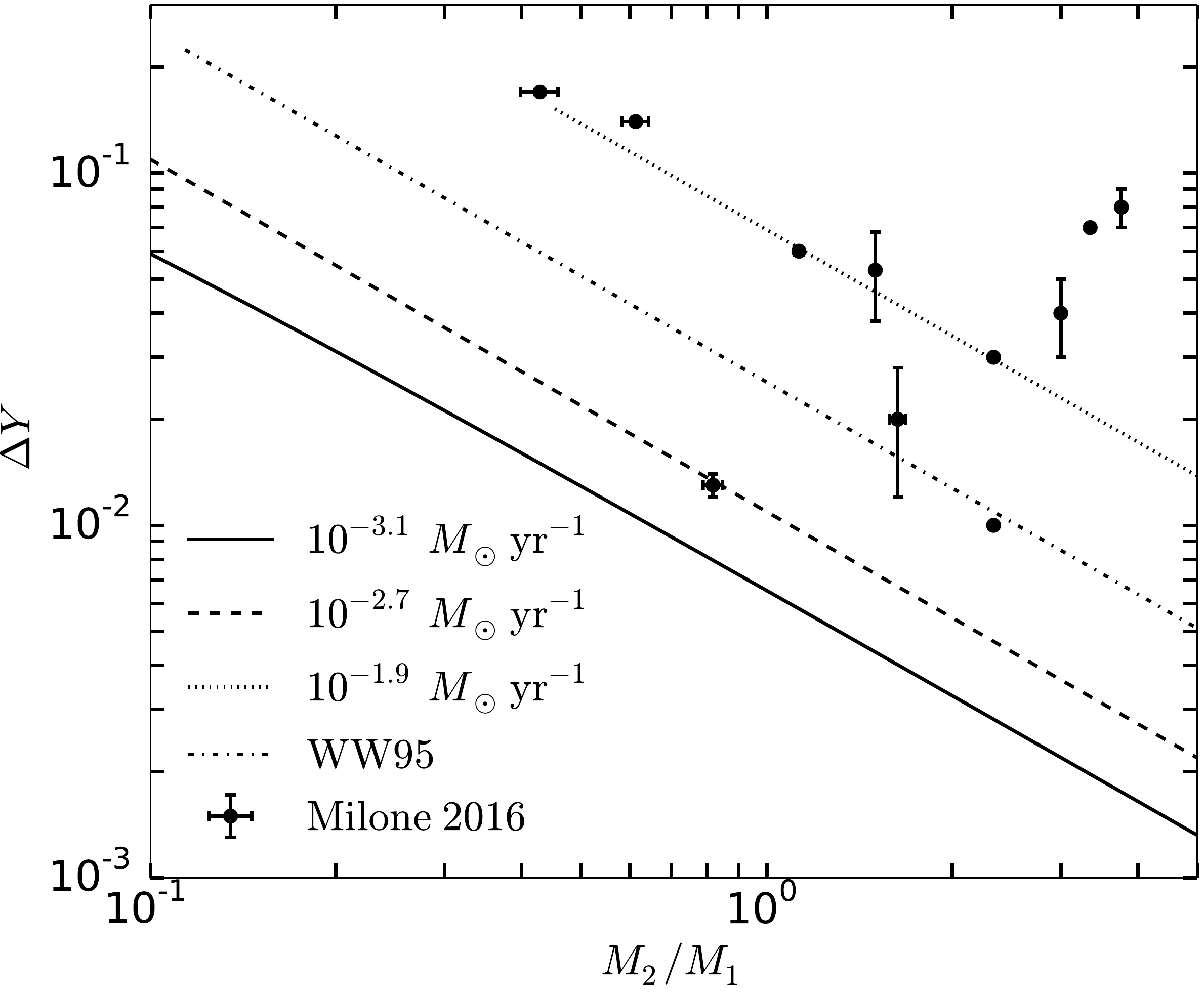}
\caption{The helium enrichment of second generation stars relative to first generation stars as a function of the ratio of second generation mass to first generation mass (equation~\ref{eq:delta_Y_scaled}). The solid line gives the relation for our fiducial cluster with $\beta_\mathrm{wind}=1$, $\dot M_w=10^{-3.1}\ M_\odot$ yr$^{-1}$, $M_1=10^5 M_\odot$, $R_{cl}=1$ pc, $Z=0.1Z_\odot$, $\alpha=0.1$, and $\Delta t_\mathrm{SN}=4.5$ Myr. The dashed line gives the relation for a similar cluster with $\beta_\mathrm{wind}\sim2.5$, which gives $\dot M_\mathrm{max}=10^{-2.7}\ M_\odot$ yr$^{-1}$. The black points are inferred values of $\Delta Y$ and measured $M_2/M_1$ for a handful of clusters presented in \citet{Milone2015a} (see their Figure 10), the dotted line is our relation for a cluster with $\dot M=10^{-1.9}\ M_\odot$ yr$^{-1}$, and the dot-dashed line is the relation for a cluster with He ejecta values from \citet{Woosley1995}. Typical quoted errors on $M_2/M_1$ are $\sim3\%$, and we plot values of $\Delta Y$ with their quoted errors. Note the dotted line is not a fit to the data.}
\label{fig:He}
\end{figure}

If the wind material does not mix with any ambient gas, $\Delta Y$ for the second generation is equal to that of stellar winds alone. The winds have a very high helium enhancement, $\Delta Y\approx0.16$, higher than most observed GCs. $\omega$ Cen has an anomalously high helium enrichment value typically not seen in other GCs, $\Delta Y=0.14$ \citep{Piotto2005}. However, without mixing with ambient gas, a second generation with mass comparable to the first generation cannot form. In addition, dilution with pristine, ambient gas is required to reproduce observed elemental abundance variations \citep[][see \S\ref{sec:other_abundances} below]{DErcole2011} and lithium content \citep{Prantzos2006}. We can produce somewhat higher second generation helium enhancements by enhancing the wind mass loss rates over our reference value of $\dot M_w$ (equation~\ref{eq:Mdot}), but still require some mixing with ambient gas. For example, if $\beta_\mathrm{wind}\sim2$, making $\dot M_w$ 2 times larger than our reference value in equation~(\ref{eq:Mdot}), then the winds would need to mix with $M_\mathrm{gas}/M_1\sim0.76$ to produce $\Delta Y\sim0.02$ (a typical value of $\Delta Y$ for MW GCs) \emph{and} also produce a second generation of size $M_2/M_1\sim0.6$, consistent with the lowest observed values of $M_2/M_1$.

We can estimate an upper bound on the mass of wind material deposited by post-main sequence massive stars by assuming that all mass lost between the zero-age main sequence (ZAMS) mass of the star and its pre-supernova mass is deposited as winds during $\Delta t_\mathrm{SN}\approx4.5$ Myr. Since most of a massive star's wind material will be deposited near the end of its lifetime, those stars with masses less than $25M_\odot$ (the lowest mass star that does not collapse directly to black hole in our model) will not deposit all of their wind mass during $\Delta t_\mathrm{SN}$, but assuming that they do provides an effective upper limit on $\beta_\mathrm{wind}$, and thus $\dot M_w$. We calculate this total mass loss as the difference between the ZAMS mass and final pre-supernova mass from Table 2 of \linebreak \citet{Sukhbold2016}, which gives the pre-supernova mass of all stars in the $9-120\ M_\odot$ range. Dividing this sum of material by $\Delta t_\mathrm{SN}=4.5$ Myr, the time between massive star winds ``turn on" and the first supernova, when we assume winds are most active, gives a value for a maximum mass deposition rate of
\begin{equation}
\dot M_{w,\mathrm{max}}=10^{-2.7}\ M_\odot\ \mathrm{yr}^{-1}\ \left(\frac{M_1}{10^5\ M_\odot}\right), \label{eq:Mdotmax}
\end{equation}
which is $\sim2.5$ times higher than our reference value of $\dot M_w$ in equation~(\ref{eq:Mdot}), implying $\beta_\mathrm{wind,max}\sim2.5$.\footnote{Our $\dot M_{w,\mathrm{max}}$ is $\sim0.2$ dex lower than the implied mass loss rate from \citet{deMink2009} for massive binaries, if we assume their reported total wind mass of 13\% of the first generation stellar mass is entirely deposited during $\Delta t_\mathrm{SN}$, before supernovae. Our value is lower because we calculate the wind mass as the difference between the initial stellar mass and the pre-supernova stellar mass, and \citet{deMink2009} calculate the difference between the initial stellar mass and the (post-supernova) remnant mass.}

For $\dot M_{w,\mathrm{max}}$, $\beta_\mathrm{crit,max}\sim0.8$, so no mass loading at all is necessary for a significant volume of the wind material to cool. However, producing a second generation mass $\sim0.5-3$ times the first generation requires a mass-loading value of $\beta\sim7-40$, compared to $\beta\sim20-100$ for our reference value of $\dot M_w$ in equation~(\ref{eq:Mdot}). Because $\dot M_{w,\mathrm{max}}$ is an estimate for the maximum mass deposition the wind can have, we interpret additional mass loading as mixing with ambient gas. Mass-loading of $\beta\sim7-40$ corresponds to masses of ambient gas $M_\mathrm{gas}/M_1\sim0.5-3.7$, compared to $M_\mathrm{gas}/M_1\sim0.5-4$ when $\beta_\mathrm{wind}=1$.

Increasing $\dot M_w$ to $\dot M_{w,\mathrm{max}}$ has a small effect on the mass of ambient gas required because the amount of total mass in the second generation is dominated by ambient gas mass. However, there is a strong effect on abundances. Figure~\ref{fig:He} shows $\Delta Y$ for models with $\dot M_\mathrm{max}$ as a function of $M_2/M_1$ as the dashed line. Increasing $\dot M_w$ to $\dot M_\mathrm{max}$ changes $\Delta Y$ from $\sim0.002-0.013$ in our fiducial model to $\Delta Y\sim0.004-0.023$. Since no ambient gas mixing is necessary for cooling for $\dot M_\mathrm{max}$, $\Delta Y$ could be as high as $Y_\mathrm{wind}-Y_1\approx0.16$ if the wind material does not mix with any ambient gas, but without any mass loading, the wind alone would only form a second generation of size $M_2/M_1\sim0.06$. To produce a helium enhancement of $\Delta Y\sim0.02$, the winds could mix with an amount of ambient gas $M_\mathrm{gas}/M_1\sim0.66$, which would produce a second generation of size $M_2/M_1\sim0.6$. Even $\dot M_{w,\mathrm{max}}$ cannot produce both helium enhancements larger than $\Delta Y\sim0.025$ \emph{and} a second generation larger than $M_2/M_1\sim0.5$.

We make a final estimate for the maximum amount of helium that could in principle be ejected before supernovae set in by imagining that all stars lose their hydrogen and helium envelopes before explosion. We estimate the helium envelope mass from the models of \citet{Woosley1995}, who calculated the evolution and explosion of massive stars. Roughly $\sim1/3$ of the ZAMS mass is contained in the helium shell and $\sim1/3$ in the hydrogen shell. Integrating over the IMF, we find a maximum allowed mass lost in winds if all massive stars eventually collapse as bare cores, with successful explosions producing Type Ic supernovae. Assuming all of this mass is ejected during $\Delta t_\mathrm{SN}$ produces a mass loss rate of $\dot M_\mathrm{WW95}\sim10^{-2.5}\ M_\odot$ yr$^{-1}$ for a cluster of mass $10^5\ M_\odot$. The dot-dashed line in Figure~\ref{fig:He} shows the relation between $\Delta Y$ and $M_2/M_1$ resulting from a wind mass loss rate of $\dot M_\mathrm{WW95}\sim10^{-2.5}\ M_\odot$ yr$^{-1}$ where half of the ejected wind mass is helium. This relation still underpredicts the inferred $\Delta Y$ and $M_2/M_1$ of observed GCs.

To reproduce the GC $\Delta Y$ data, our model would need a wind mass loss rate of $\dot M\sim10^{-2}\ M_\odot$ yr$^{-1}$ (shown as dotted line in Figure~\ref{fig:He}), larger than our simple estimate for the maximum mass loss rate a GC could have. Such a mass-loss rate would imply 45\% of the first stellar generation's mass is ejected in winds, $\sim12$ times larger than our fiducial $\dot M_w$, $\sim5$ times larger than $\dot M_{w,\mathrm{max}}$ in equation~(\ref{eq:Mdotmax}), and $\sim3$ times larger than $\dot M_{WW}$ derived above, and would require a flatter high-mass IMF than we have assumed. This emphasizes the problem of all self-enrichment models, including our own, where there is not enough wind material available to produce a large enough second generation with large enough elemental enhancements. We emphasize that while the helium enhancement \citep{Milone2015a} and second generation fraction \citep{Milone2017} are both seen to increase with GC mass, we do not find a strong correlation or anti-correlation between $\Delta Y$ and $M_2/M_1$ for clusters reported in the literature. If such an anti-correlation is established, it would provide evidence for the self-enrichment scenario. For all but one of the observed GCs plotted in Figure~\ref{fig:He}, our models, including the one with a maximum mass loss rate that we estimate a GC could have, underestimate either the reported $\Delta Y$ values or the reported $M_2/M_1$ values for literature GCs.

We can calculate a $\Delta Z$ analogous to our calculation of $\Delta Y$, where instead of using the helium mass content of the wind material, we sum the mass content of all elements heavier than helium in the wind. Regardless of the fraction of wind material in the mixture of ambient gas and wind material, $\Delta Y/\Delta Z\sim1.6$ in our model. This does not depend on $M_\mathrm{gas}$ because $\Delta Y$ and $\Delta Z$ depend on $M_\mathrm{gas}$ in the same way, so all $M_\mathrm{gas}$ dependence drops out in the ratio $\Delta Y/\Delta Z$.

Because the time for the second generation of stars to form, $\Delta t_\mathrm{2gen}=t_\mathrm{ff}+t_\mathrm{cool}$ (see Figure~\ref{fig:tff}), is shorter than the time $\Delta t_\mathrm{SN}$ when the first supernova that does not collapse directly to black hole occurs, it is possible for the second generation to form before massive star winds have deposited very much of their mass into the cluster. We assumed above that all of the wind material massive stars deposit over their entire post-main sequence evolution enters into the second stellar generation, when more precisely the amount of wind material in the second generation would be given by $M_w=\dot M_w \Delta t_{2gen}$ when $\beta_\mathrm{wind}=1$. This would create a second generation that is even more diluted by ambient gas than what we derive above, because $\Delta t_\mathrm{2gen}\lesssim\Delta t_\mathrm{SN}$. Even under the assumption that $\Delta t_\mathrm{2gen}\sim\Delta t_\mathrm{SN}$, the second generation mass is dominated by the ambient gas mass, so $\Delta t_\mathrm{2gen}<\Delta t_\mathrm{SN}$ does not have a large impact on the amount of ambient gas required to form a large second generation --- $M_\mathrm{gas}/M_1\sim1.24$ to produce $M_2/M_1\sim1$ for $\Delta t_\mathrm{2gen}\sim1$ Myr, compared to $M_\mathrm{gas}/M_1\sim1.21$ to produce the same $M_2/M_1$ under the assumption $\Delta t_\mathrm{2gen}\sim\Delta t_\mathrm{SN}\sim4.5$ Myr. However, the difference in the value of $\beta$ required to produce $M_2/M_1\sim1$ is quite large, $\beta\sim157$ for $\Delta t_\mathrm{2gen}\sim1$ Myr, compared to $\beta\sim35$ in our fiducial model, which means the difference in $\Delta Y$ is also large: $\Delta Y\sim0.001$ for $\Delta t_\mathrm{2gen}\sim1$ Myr compared to $\Delta Y\sim0.005$ for our fiducial model. Shortening $\Delta t_\mathrm{2gen}$ thus has a large impact on the second generation abundances, but not on the mass of ambient gas required to produce a large second generation. Because $\Delta t_\mathrm{2gen}$ and $\Delta t_\mathrm{SN}$ are of the same order of magnitude, we scale $\Delta t_\mathrm{2gen}$ to $\Delta t_\mathrm{SN}$ as a reference, but note that the precise value of $\Delta t_\mathrm{2gen}$ impacts $\Delta Y$.

%%%%%%%%%%%%%%%%%%%%%%%%%%%%%%%%%%%%%%%%%%%%%%%%%%%%%%%%%%%%%%%%%%%%%%%%%%%%%%%%%%%%%%%%%
\subsection{Other Light Element Abundances}
\label{sec:other_abundances}
%%%%%%%%%%%%%%%%%%%%%%%%%%%%%%%%%%%%%%%%%%%%%%%%%%%%%%%%%%%%%%%%%%%%%%%%%%%%%%%%%%%%%%%%%

In observed GCs, the variations in light element abundances determine which stars are classified as members of the first or second generations. The most notable light-element spread is the sodium-oxygen anticorrelation, in which second generation stars are Na-enriched and O-poor, whereas first-generation stars are O-enriched and Na-poor. The $\sim1235$ red giant stars across the 19 GCs in the C09 sample have a spread from [O/Fe]$\sim0.0-0.7$ and [Na/Fe]$\sim-0.4-0.0$ for the first generation and [O/Fe]$\sim-0.7-0.5$ and [Na/Fe]$\sim0.0-1.0$ for the second generation across all GCs studied. The iron abundances are equivalent across generations for most observed GCs, so we will focus on [O/Na]. For the first generation, [O/Na]$\sim0.0-1.0$ and for the second generation, [O/Na]$\sim-1.5-0.5$.

We analyzed the oxygen and sodium yields for the wind material in the \citet{Sukhbold2016} models, and found that the winds of stars in the mass range $30\ M_\odot < M < 120\ M_\odot$ have spreads of [O/Na]$\sim-0.2-0.4$. Because our model assumes complete mixing of winds from all stellar masses, there is no spread in abundances in the second generation. However, we can approximate a type of incomplete mixing if we instead assume winds from the most massive stars form the first of the second generation stars and winds from the least massive stars form the last of the second generation. There would then be a spread in abundances within the second stellar generation because winds from different mass stars have different abundances. Note that this method of producing a spread in abundances still will not produce as large of an abundance spread as allowing wind material to be diluted to various degrees with ambient gas. If the second generation of stars is entirely made out of wind material, with no mixing with pristine ambient gas at all, then the wind abundances would exactly represent the abundances for the second generation in our model (but the second generation would be small). The spread in [O/Na] is not as wide as the observed spread in second generations, because we do not include variable mixing of wind material with ambient gas.

To create a second stellar generation $0.5-3$ times more massive than the first, mixing with a mass of ambient gas $M_\mathrm{gas}/M_1\sim0.5-4$ is necessary for our reference value of $\dot M_w$ ($\beta_\mathrm{wind}=1$). We assume the ambient gas and first stellar generation have one-tenth solar Fe abundances, and use [O/Fe]$\sim0.35$ and [Na/Fe]$\sim-0.1$ for the ambient gas, to match the oxygen and sodium abundances of the first stellar generation. After mixing, the wind abundances are diluted by the ambient gas, so the spread in abundances of the second generation becomes [O/Na]$\sim-0.1-0.4$ for $M_\mathrm{gas}/M_1\sim0.5$ and [O/Na]$\sim0.18-0.45$ for $M_\mathrm{gas}/M_1\sim4$. When mixing with such a large mass of ambient gas, the second generation abundances do not match the spread of the observed second generation abundances, but the values of [O/Na] are still consistent with the second generation.

If we increase $\dot M_w$ to $\dot M_\mathrm{max}=10^{-2.7}\ M_\odot$ yr$^{-1}$, a somewhat smaller value of $M_\mathrm{gas}/M_1\sim0.5-3.7$ is necessary to produce a second generation ratio $M_2/M_1\sim0.5-3$. The value of [O/Na] for the winds alone without mixing are the same for $\dot M_\mathrm{max}$ as they are for $\dot M_w$ because increasing the overall magnitude of the wind does not change its relative abundances. However, using $\dot M_\mathrm{max}$ does change the relative amounts of wind material and ambient gas in the mixture, so the spreads in [O/Na] are different than for $\dot M_w$ when mixing with various amounts of ambient gas. For $M_\mathrm{gas}/M_1\sim0.5$, [O/Na]$\sim-0.15-0.42$, and for $M_\mathrm{gas}/M_1\sim3.7$, [O/Na]$\sim0.04-0.43$. Again, the [O/Na] values do not match the full spread observed in GCs, but they are consistent with the observed second generation [O/Na]. A wind model that allows varying degrees of mixing between wind material and ambient gas is necessary to produce the full range of observed [O/Na] values \citep{Prantzos2006}.

%%%%%%%%%%%%%%%%%%%%%%%%%%%%%%%%%%%%%%%%%%%%%%%%%%%%%%%%%%%%%%%%%%%%%%%%%%%%%%%%%%%%%%%%%
\section{Discussion}
\label{sec:discussion}
%%%%%%%%%%%%%%%%%%%%%%%%%%%%%%%%%%%%%%%%%%%%%%%%%%%%%%%%%%%%%%%%%%%%%%%%%%%%%%%%%%%%%%%%%

Throughout this paper, we have assumed a fiducial model for GCs and developed analytic criteria for the radiative cooling of mixed wind material and ambient gas to promote formation of a second generation of stars. We have assumed a metallicity of one-tenth solar, no feedback processes other than energy and mass deposition of winds from massive, non-rotating stars, and only a single star formation event after the initial formation of the cluster. In this section, we examine how adjustments to our assumptions would impact our results.

%%%%%%%%%%%%%%%%%%%%%%%%%%%%%%%%%%%%%%%%%%%%%%%%%%%%%%%%%%%%%%%%%%%%%%%%%%%%%%%%%%%%%%%%%
\subsection{Very Low Metallicities}
%%%%%%%%%%%%%%%%%%%%%%%%%%%%%%%%%%%%%%%%%%%%%%%%%%%%%%%%%%%%%%%%%%%%%%%%%%%%%%%%%%%%%%%%%

The value of $\beta_\mathrm{crit}$ derived in equation~(\ref{eq:beta_crit}) is dependent on the cooling function, which depends on the metallicity and relative metal abundances. Perhaps counter-intuitively, decreasing the metallicity of the first generation decreases $\beta_\mathrm{crit}$ because of changes in $\dot E_w$ and $\dot M_w$, despite the decrease in the normalization of the cooling function. The fraction of the mixture that is made up of the wind material affects the overall gas metallicity, since massive star winds are enriched in certain H-burning byproducts, such as sodium and aluminum, and depleted in other elements such as oxygen and magnesium \citep{Gratton2012}. Indeed, we find that the metallicity of all metals in just the wind material in our fiducial cluster is about half that of solar, despite the first generation stars having one-tenth solar metallicity. This means that the critical mass-loading $\beta_\mathrm{crit}$ is actually dependent on the mass-loading itself. We have avoided this dependence in our analytic formulation by assuming a metallicity and checking that the mixture of wind material and ambient gas that we require has a metallicity somewhat similar to our assumption.

For example, using \citet{Sukhbold2016}, winds from first-generation stars formed with one-tenth solar metallicity mixing with ambient gas of one-tenth solar metallicity, for our fiducial values of $M_\mathrm{gas}/M_1\sim0.5-4$ when $\beta_\mathrm{wind}=1$, the mixture has a metallicity of $\sim0.11$ solar due to the high metallicity of the wind. We assume this to be close enough to the value of $Z_\mathrm{gas}=0.1Z_\odot$ that we use for our analytic prescription. A more thorough analysis would involve tracking the wind metal species and adjusting the cooling function accordingly.

%%%%%%%%%%%%%%%%%%%%%%%%%%%%%%%%%%%%%%%%%%%%%%%%%%%%%%%%%%%%%%%%%%%%%%%%%%%%%%%%%%%%%%%%%
\subsection{Other Feedback Processes}
%%%%%%%%%%%%%%%%%%%%%%%%%%%%%%%%%%%%%%%%%%%%%%%%%%%%%%%%%%%%%%%%%%%%%%%%%%%%%%%%%%%%%%%%%

We have assumed that only the mass and energy of massive star winds contribute to the feedback processes in the GC. Theoretical and observational studies by, e.g., \citet{Murray2010,Lopez2011,Murray2011} show that radiation pressure on dust may blow gas out of GCs, and disrupt the host giant molecular cloud, on timescales less than a few Myr, before the first supernovae, potentially calling into question our assumption of a large reservoir of gas remaining for $\sim$Myr timescales after formation of the first stellar generation. However, the prevalence of dust in the mixture of gas in GCs likely depends on the metallicity of the gas; more dust forms at higher metallicities. Since radiation pressure is more effective at higher dust-to-gas ratios, higher metallicity clusters may inhibit second generation star formation by rapidly expelling gas. Strong radiation pressure feedback in high metallicity clusters may thus make low-metallicity GCs relatively more efficient at forming a second generation. An interesting avenue for further investigation would be an analysis of all feedback processes in GCs and how they relate to secondary star formation in a time dependent model.

%%%%%%%%%%%%%%%%%%%%%%%%%%%%%%%%%%%%%%%%%%%%%%%%%%%%%%%%%%%%%%%%%%%%%%%%%%%%%%%%%%%%%%%%%
\subsection{Asymptotic Giant Branch Stellar Winds}
\label{sec:AGB}
%%%%%%%%%%%%%%%%%%%%%%%%%%%%%%%%%%%%%%%%%%%%%%%%%%%%%%%%%%%%%%%%%%%%%%%%%%%%%%%%%%%%%%%%%

AGB winds have been examined as a potential source of material to form a second stellar generation \citep{Ventura2001,Conroy2011,Conroy2012} --- like massive star winds, they are enriched in light element byproducts of hydrogen burning, and their low velocities allow the GC to retain them without requiring kinetic energy be lost to radiative cooling. \citet{Conroy2015} provide
\begin{align}
\dot M_\mathrm{AGB}&\approx 10^{-4.1}\ M_\odot\ \mathrm{yr}^{-1}\ \left(\frac{M_1}{10^5\ M_\odot}\right)\left(\frac{t_{cl}}{10^8\ \mathrm{yr}}\right)^{-1.25} \\
\dot E_\mathrm{AGB}&\approx 10^{33.9}\ \mathrm{erg\ s}^{-1}\ \nonumber \\
&\times\left(\frac{M_1}{10^5\ M_\odot}\right)\left(\frac{\sigma}{10\ \mathrm{km\ s}^{-1}}\right)^2\left(\frac{t_{cl}}{10^8\ \mathrm{yr}}\right)^{-1.25}
\end{align}
where $t_{cl}$ is the age of the stellar cluster when AGB winds are active, $\sigma$ is the velocity dispersion of the GC, and we have scaled $\dot M_\mathrm{AGB}$ and $\dot E_\mathrm{AGB}$ to our fiducial GC parameters and typical parameters for $\sigma$ \citep{McLaughlin2005} and timescales for AGB winds. Substituting $\dot M_\mathrm{AGB}$ and $\dot E_\mathrm{AGB}$ into equation~(\ref{eq:beta_crit}) gives $\beta_\mathrm{crit,AGB}\sim0.03$ and $\beta_\mathrm{crit,min,AGB}\sim0.01$ for $\alpha=0.1$, and $\beta_\mathrm{crit,AGB}\sim0.15$ and $\beta_\mathrm{crit,min,AGB}\sim0.07$ for $\alpha=1$. The critical condition on $M_1/R_{cl}$ for AGB winds becomes $[M_1/R_{cl}]_\mathrm{crit,AGB}\sim10^3\alpha_{0.1}^{2.7}\beta^{-3.7}\ M_\odot$ pc$^{-1}$. Since the necessary mass-loading for AGB winds to cool is very small, the winds will easily be able to cool without requiring additional mass-loading by ambient gas. However, the AGB winds scenario suffers from the same problems as the massive star winds scenario. In order to form a second generation with a half to three times as much mass as the first, assuming $\beta_\mathrm{wind}=1$, 
\begin{align}
&\beta_\mathrm{AGB}=\frac{M_2}{\dot M_\mathrm{AGB}(R_\mathrm{cool}/R_{cl})^3\Delta t_\mathrm{AGB}} \\
&\sim 20\left(\frac{M_2}{2.5M_1}\right)\left(\frac{R_\mathrm{cool}}{0.93R_{cl}}\right)^{-3}\left(\frac{\Delta t_\mathrm{AGB}}{10^8\ \mathrm{yr}}\right)^{-1} \nonumber \\
&\times\left(\frac{\dot M_\mathrm{AGB}}{10^{-4.1}M_\odot\ \mathrm{yr}^{-1}}\right)^{-1}
\end{align}
where $\Delta t_\mathrm{AGB}\sim 10^8$ yr is the length of time over which AGB winds are active, and we take $M_2=2.5M_1$ and $(R_\mathrm{cool}/R_{cl})^3=0.8$. $\beta_\mathrm{AGB}\sim4-20$ corresponds to a mass of ambient gas of $M_\mathrm{gas}\sim0.5-5M_1$ (see equation~\ref{eq:Mgas}), similar to the $M_\mathrm{gas}$ required by the massive star winds scenario when $\beta_\mathrm{wind}=1$. Unlike the massive star winds scenario, a high mass loading in the AGB winds scenario requires accretion of gas a hundred Myr after GC formation instead of mixing with leftover gas from the first generation of star formation \citep[for a discussion of GC gas accretion scenarios, see e.g.][]{Conroy2011}. Both the massive star winds and the AGB winds scenarios may operate for a GC \citep[][find evidence for the AGB scenario in NGC 2808]{DAntona2016}.

%%%%%%%%%%%%%%%%%%%%%%%%%%%%%%%%%%%%%%%%%%%%%%%%%%%%%%%%%%%%%%%%%%%%%%%%%%%%%%%%%%%%%%%%%
\subsection{Rotating Star Wind Models}
%%%%%%%%%%%%%%%%%%%%%%%%%%%%%%%%%%%%%%%%%%%%%%%%%%%%%%%%%%%%%%%%%%%%%%%%%%%%%%%%%%%%%%%%%

We examined the $\dot E_w$ and $\dot M_w$ used by STARBURST99 in the rotating models for one-tenth solar metallicity, in which stars rotate at $0.4$ of break-up, originally presented in \citet{Georgy2013}. $\dot M_w$ and $\dot E_w$ do not change significantly when stellar rotation is included, so the primary effect of including rotating star models is to increase the time until the first supernova that does not collapse directly to black hole by $\sim0.5$ Myr, potentially prolonging the time over which second generation star formation can occur. However, in Figure~\ref{fig:tff} we show that for much of the parameter space explored here, a second generation can form on a $\sim1$ Myr timescale, so the additional $\sim0.5$ Myr does not qualitatively change our results.

%%%%%%%%%%%%%%%%%%%%%%%%%%%%%%%%%%%%%%%%%%%%%%%%%%%%%%%%%%%%%%%%%%%%%%%%%%%%%%%%%%%%%%%%%
\subsection{More Than Two Generations}
%%%%%%%%%%%%%%%%%%%%%%%%%%%%%%%%%%%%%%%%%%%%%%%%%%%%%%%%%%%%%%%%%%%%%%%%%%%%%%%%%%%%%%%%%

Some GCs show evidence of more than two stellar populations \citep[e.g.,][]{Piotto2007,Carretta2012,Carretta2015,Milone2015b}. Since we have found the second generation of stars may form in $\sim1$ Myr after massive stars reach post-main sequence evolution, there is a possibility that more than two generations of stars may form in a single GC before supernovae pollute the gas in the GC with heavy elements $\sim3-5$ Myr later. Since $\beta_\mathrm{crit}\sim1$, radiative cooling should again occur. However, a large $\beta$ is again necessary for a substantial population to form. The wind material from massive second generation stars will be even more enriched with light element products of hydrogen burning than that from the first generation, or potentially AGB winds (see \S\ref{sec:AGB}).

%%%%%%%%%%%%%%%%%%%%%%%%%%%%%%%%%%%%%%%%%%%%%%%%%%%%%%%%%%%%%%%%%%%%%%%%%%%%%%%%%%%%%%%%%
\section{Conclusions}
\label{sec:conclusions}
%%%%%%%%%%%%%%%%%%%%%%%%%%%%%%%%%%%%%%%%%%%%%%%%%%%%%%%%%%%%%%%%%%%%%%%%%%%%%%%%%%%%%%%%%

We have explored a model where winds from massive stars in globular clusters collide with each other and thermalize, then rapidly radiatively cool, forming a second generation of stars before the first supernovae \citep[see also][]{Wunsch2008,Palous2014,Wunsch2017}. We use the mass and energy deposition of massive stars and examine only winds that are produced before the first supernova that does not collapse to a black hole, because we find the second stellar generation can form well before supernovae are expected to occur. Our main results are:
\begin{enumerate}
\item The minimum mass-loading for any of the cluster's gas to cool and be retained within the cluster is given in equation~(\ref{eq:beta_crit_min}). Thus, for $\beta\sim1$, $\alpha\sim0.1$, and $M_1/R_{cl}\gtrsim10^5\ M_\odot$ pc$^{-1}$, no additional mass other than that provided by massive star winds is necessary for the winds to radiatively cool.
\item The mass loading required for nearly all of the cluster to cool is just $\sim2$ times higher than the minimum mass-loading for any of the cluster to cool (equation~\ref{eq:beta_crit_scaled}).
\item For $\beta>\beta_\mathrm{crit}$, the cluster wind's momentum and energy deposition rates ($\dot p_w$ given by equation~\ref{eq:pdot_esc} and $\dot E_w$ given by equation~\ref{eq:Edot_esc}) rapidly decrease as $\beta$ increases. We speculate that a reduced cluster wind allows gas to remain within or in proximity to the cluster, and predict that low mass clusters should not have a second stellar generation (Figure~\ref{fig:MdRvsM}).
\item To produce a second stellar generation that is a half to three times larger than the first, a mass-loading of $\beta\sim20-100$ is necessary, and does not strongly depend on $M_1$, $R_{cl}$, or $\alpha$. A high mass-loading can be provided by mixing the mass of wind material, which may be larger than expected from current stellar models in post-main sequence evolution, with an ambient gas mass leftover from the formation of the first stellar generation. $\beta\sim20-100$ corresponds to $M_\mathrm{gas}/M_1\sim0.5-4$ if wind mass loss rates are not enhanced. To produce a second generation mass fraction of $M_2/M_1\sim1$, a mass loading factor of $\beta\sim30$, corresponding to $M_\mathrm{gas}/M_1\sim1$ in our fiducial model, is needed. The onset time of supernovae, the mass ejection by winds, and the black hole formation probability for supernovae are uncertain. If all massive stars lose all of their pre-supernova mass as winds before the first supernovae, they drive a maximum mass-loss rate of $\dot M_\mathrm{max}\sim10^{-2.7}\ M_\odot$ yr$^{-1}$. For $\dot M_\mathrm{max}$, $\beta_\mathrm{crit}\sim0.8$ and the mass loading required to produce a second generation of size $M_2/M_1\sim0.5-3$ is $\beta\sim7-40$, corresponding to $M_\mathrm{gas}/M_1\sim0.5-3.7$.
\item The sum of the cooling and free-fall times for mass-loaded massive star winds are small, $\sim1-2$ Myr, showing that the second generation can form before the first supernovae pollute the cluster gas with iron and other heavy elements.
\item If winds mix with an amount of gas $M_\mathrm{gas}/M_1\sim0.5-4$, the helium enhancement in the second generation is $\Delta Y\sim 0.002-0.013$, smaller than most observed helium spreads in globular clusters with multiple generations (see Figure~\ref{fig:He}). However, if high mass-loading is provided by enhancing massive star winds and mixing with ambient gas, the second generation helium enhancement is $\Delta Y\sim0.004-0.023$ (equation~\ref{eq:delta_Y_scaled}, dashed line in Figure~\ref{fig:He}) or $\Delta Y\sim0.008-0.05$ for extreme assumptions about helium mass loss (dot-dashed line in Figure~\ref{fig:He}). He enhancement up to $\Delta Y\sim0.16$ is possible if wind material does not mix with ambient gas at all, but this yields a smaller second generation fraction than observed values (see Figure~\ref{fig:He}). The values of [O/Na]$\sim-0.2-0.4$ for pure wind with no mixing and [O/Na]$\sim-0.1-0.16$ for mixing with maximal ambient gas mass $M_\mathrm{gas}/M_1\sim4$ are consistent with the observed range of [O/Na]$\sim-1.5-0.5$ in second generations, but the range of [O/Na] values produced is smaller in our model because we assume total mixing of wind material with ambient gas before any second generation stars are formed.
\end{enumerate}

Rapid radiative cooling solves the problem of gas retention by GCs, and reduces the amount of extra mass needed by other self-enrichment scenarios to produce a large second generation, from $10-100$ times more mass \citep{Conroy2012} down to $0.5-4$ times more mass (although, only without fully reproducing observed second generation elemental abundances). Additionally, we do not require the extra mass needed to be in the form of first-generation stars, so our model does not require very large fractions \citep[e.g. 96\% in][]{Decressin2007b} of first-generation stars to be preferentially ejected over second-generation stars, or very flat IMFs. The model predicts that $\Delta Y$ is inversely proportional to the second generation mass fraction (equation~\ref{eq:delta_Y_scaled} and Figure~\ref{fig:He}) because mixing with ambient gas dilutes the enriched winds. Like other self-enrichment studies, our model cannot reproduce the abundance variations observed in GCs while simultaneously reproducing a massive second generation with a standard IMF, even under extreme assumptions about the total mass ejected by massive stars. A more complete understanding of the abundances and mass-loss rates of massive star winds may help to further reduce the problems of the many self-enrichment scenarios for second generation star formation.

%%%%%%%%%%%%%%%%%%%%%%%%%%%%%%%%%%%%%%%%%%%%%%%%%%%%%%%%%%%%%%%%%%%%%%%%%%%%%%%%%%%%%%%%%
\section*{Acknowledgments}
%%%%%%%%%%%%%%%%%%%%%%%%%%%%%%%%%%%%%%%%%%%%%%%%%%%%%%%%%%%%%%%%%%%%%%%%%%%%%%%%%%%%%%%%%
This work is supported in part by NSF grant 1516967. TAT thanks Dong Zhang for collaboration and discussion. CL thanks Jennifer A. Johnson, David H. Weinberg, and Laura A. Lopez for discussion of stellar feedback, and Tuguldur Sukhbold for discussion of wind abundances. We thank N. Bastian, E. Carretta, R. W{\"u}nsch, J. Palou{\v s}, G. Tenorio-Tagle, and S. Silich for constructive comments. We thank the anonymous referee for a timely and helpful report that improved the manuscript.

\end{document}